\documentclass[a4paper,11pt,fleqn]{article}
\usepackage[utf8]{inputenc}
\usepackage{amsmath}
\usepackage{amssymb}
\usepackage{mathtools}
\usepackage{empheq}
\usepackage[margin=1in]{geometry}
\usepackage{newtxmath}
\usepackage{newtxtext}
\usepackage{microtype}
\usepackage{array}
\usepackage{hhline}
\usepackage{booktabs}
\usepackage{colortbl}
\usepackage{multirow}
\usepackage{dsfont}
\usepackage{mathrsfs}
\usepackage{pdflscape}
\usepackage{longtable}
\usepackage{xspace}
\usepackage{nicefrac}
\usepackage{cite}  
\usepackage[small]{caption2}
\usepackage{nccfoots}
\usepackage{tikz}
\usetikzlibrary{calc,positioning}
\usetikzlibrary{arrows,shapes}
\usetikzlibrary{arrows.meta}
\pgfdeclarelayer{background layer}
\pgfdeclarelayer{foreground layer}
\pgfsetlayers{background layer,main,foreground layer}
\definecolor{gr0}{RGB}{235,235,235}
\definecolor{c1}{RGB}{0,119,187}
\definecolor{c2}{RGB}{51,187,238}
\definecolor{c3}{RGB}{0,153,136}
\definecolor{c4}{RGB}{238,119,51}
\definecolor{c5}{RGB}{204,51,17}
\definecolor{c6}{RGB}{238,51,119}
\definecolor{c0}{RGB}{187,187,187}
\definecolor{darkgreen}{HTML}{109930}
\definecolor{pink}{rgb}{0.858, 0.188, 0.478}
\usepackage{enumitem}
\usepackage{blkarray, bigstrut}
\usepackage{braket}
\allowdisplaybreaks[3]

\usepackage{frcursive}

\usepackage[nolist,nohyperlinks,printonlyreused]{acronym} 

\makeatletter
\newcommand{\dd}{\mathop{}\!\mathrm{d}}%
\newcommand{\ii}{\mathop{}\!\mathrm{i}\!\mathop{}}%
\newcommand{\ee}{\mathrm{e}}%
\g@addto@macro\bfseries{\boldmath}
\newcommand*{\defeq}{\mathchoice{\mathrel{\rlap{%
                     \raisebox{0.24ex}{$\m@th\cdot$}}%
                     \raisebox{-0.24ex}{$\m@th\cdot$}}%
                     =}{\mathrel{\rlap{%
                     \raisebox{0.24ex}{$\m@th\cdot$}}%
                     \raisebox{-0.24ex}{$\m@th\cdot$}}%
                     =}{\mathrel{\rlap{%
                     \raisebox{0.08ex}{\small$\m@th\cdot$}}%
                     \raisebox{-0.28ex}{\small$\m@th\cdot$}}%
                     =}{\mathrel{\rlap{%
                     \raisebox{0.08ex}{\tiny$\m@th\cdot$}}%
                     \raisebox{-0.28ex}{\tiny$\m@th\cdot$}}%
                     =}}
\newcommand*{\eqdef}{\mathchoice{=\mathrel{\rlap{%
                     \raisebox{0.24ex}{$\m@th\cdot$}}%
                     \raisebox{-0.24ex}{$\m@th\cdot$}}}{%
					 =\mathrel{\rlap{%
                     \raisebox{0.24ex}{$\m@th\cdot$}}%
                     \raisebox{-0.24ex}{$\m@th\cdot$}}}{%
					 =\mathrel{\rlap{%
                     \raisebox{0.08ex}{\small$\m@th\cdot$}}%
                     \raisebox{-0.28ex}{\small$\m@th\cdot$}}}{%
					 =\mathrel{\rlap{%
                     \raisebox{0.08ex}{\tiny$\m@th\cdot$}}%
                     \raisebox{-0.28ex}{\tiny$\m@th\cdot$}}}%
                     }
\makeatother

\DeclareMathOperator{\im}{Im}

\newcommand{\rep}[1]{\ensuremath\boldsymbol{#1}}

\newcommand{\Z}[1]{\ensuremath{\mathds{Z}_{#1}}} 

\newcommand{\SL}[1]{\ensuremath{\text{SL}(#1)}}

\newcommand{\Id}{\mathds{1}}

\newcommand{\CP}{\ensuremath{\mathcal{CP}}\xspace}

\newcommand{\vev}[1]{\ensuremath{\langle{#1}\rangle}}

\DeclareMathOperator{\image}{im}

\usepackage[pdftex]{hyperref}
\hypersetup{%
    colorlinks = true,
    linkcolor = blue,
    anchorcolor = blue,
    citecolor = blue,
    filecolor = blue,
    urlcolor = blue,
	pdftitle = {Multiple realizations of modular flavor symmetries and their phenomenology},
	pdfauthor = {Arriaga-Osante, Chen, Diaz-Castro, Li, Liu, Ramos-Sanchez, Ratz}
}
\usepackage{cleveref}

\allowdisplaybreaks

\begin{document}
\colorlet{lightgray}{gr0}

	\begin{titlepage}
		
		\begin{flushright}
			\normalsize{UCI-TR 02/25}
		\end{flushright}
		
		\vspace*{1.0cm}

		\begin{center}
		    {\Large\textbf{Multiple realizations of modular flavor symmetries\\[1ex] and their phenomenology}}
			
			\vspace{1cm}
			\textbf{Carlos Arriaga-Osante}\rlap{,}$^1$
			\textbf{Mu-Chun Chen}\rlap{,}$^2$
			\textbf{Ram\'on D\'iaz-Castro}\rlap{,}$^1$
			\textbf{Xueqi Li}\rlap{,}$^2$ \par
			\textbf{Xiang-Gan Liu}\rlap{,}$^2$
			\textbf{Sa\'ul Ramos-S\'anchez}$^1$ and
			\textbf{Michael Ratz}$^2$
			\Footnote{*}{%
				\href{mailto:cebrinarriaga@ciencias.unam.mx;muchunc@uci.edu;jesusramon@estudiantes.fisica.unam.mx;xueqi.li@uci.edu;xianggal@uci.edu;ramos@fisica.unam.mx;mratz@uci.edu}{\texttt{cebrinarriaga@ciencias.unam.mx; muchunc@uci.edu; jesusramon@estudiantes.fisica.unam.mx; xueqi.li@uci.edu; xianggal@uci.edu; ramos@fisica.unam.mx; mratz@uci.edu}}
			}
			\\[5mm]
			$^1$\textit{\small Instituto de F\'isica, Universidad Nacional Aut\'onoma de M\'exico,\\ Cd.~de M\'exico C.P.~04510, M\'exico}\\
			$^2$\textit{\small Department of Physics and Astronomy, University of California, Irvine, CA 92697-4575, USA}
		\end{center}
		
		\vspace{1cm}
		
		\vspace*{1.0cm}
		
		\begin{abstract}
			We point out that specifying the finite modular group does not uniquely fix a modular flavor symmetry. We illustrate this using the finite modular group $T'$.
			Otherwise equivalent models based on different $T'$ lead to modular forms with different properties and, hence, produce different phenomenological features. 
			We exemplify this in various scenarios, and show that the ability of a given model to accommodate mass and other observed hierarchies depends sensitively on the way the $T'$ is implemented. 
		\end{abstract}
		
	\end{titlepage}
	
	\clearpage

\begin{acronym}
	\acro{SM}{Standard Model}
	\acro{CG}{Clebsch--Gordan}
	\acro{CKM}{Cabibbo--Kobayashi--Maskawa}
	\acro{irrep}{irreducible representation}
	\acro{MLDE}{modular linear differential equation}
	\acro{SUSY}{supersymmetry}
	\acro{UV}{ultraviolet}
	\acro{VEV}{vacuum expectation value}
	\acro{VVMF}{vector-valued modular form}
\end{acronym}

\section{Introduction}
\label{sec:Introduction}

The \ac{SM} describes particle physics very well, but at the expense of introducing close to 30 parameters.
These parameters have no explanation within the \ac{SM}. Most of these parameters reside in the flavor sector.
This constitutes the so-called flavor puzzle, which concerns the origin of fermion masses, mixings and hierarchies, and \CP violation.

A promising, and relatively new approach, to address the flavor problem, proposed by Feruglio~\cite{Feruglio:2017spp}, assumes that Yukawa couplings are \acp{VVMF}.
In the simplest incarnation of the resulting scheme, these \acp{VVMF} depend on a single complex modulus, $\tau$.
\acp{VVMF} are holomorphic and transform nontrivially under a finite modular group $G$.
They are incorporated in supersymmetric models, in which matter fields furnish representations under $G$.
In other words, $G$ acts as a flavor symmetry, and correspondingly the scheme has been dubbed ``modular flavor symmetries''.

One may ask whether specifying the finite modular group $G$ uniquely determines the \acp{VVMF}, and thus the predictions of the model.
The answer to this question is no~\cite{Liu:2021gwa}.
While many promising results have been obtained in scenarios in which the finite modular group $G$ is the quotient of the modular group $\SL{2,\mathds{Z}}$ and its principal congruence subgroups $\Gamma(N)$~\cite{Feruglio:2017spp,Kobayashi:2018vbk,Kobayashi:2018scp,Kobayashi:2018wkl,Novichkov:2018yse,Penedo:2018nmg,Novichkov:2018nkm,Novichkov:2018ovf,Kobayashi:2019gtp,Kobayashi:2019mna,Liu:2019khw,Ding:2019zxk,Ding:2019xna,Ding:2019gof,Feruglio:2019ybq,Criado:2019tzk,Wang:2020lxk,Ding:2020msi,Qu:2021jdy,Ding:2022nzn,Meloni:2023aru,Kikuchi:2023cap},
there are in fact additional possibilities which have gotten far less attention so far.

One can construct finite modular groups using not only the principal congruence subgroups but also other normal subgroups of $\SL{2,\mathds{Z}}$ with finite index~\cite{Liu:2021gwa}.
Once a finite modular group has been chosen as flavor symmetry, the next step to arrive at a model of flavor is to determine the modular forms of the chosen finite modular group.
Systematic methods for obtaining the \acp{VVMF} associated with these groups have been developed~\cite{bantay2007vector,Liu:2021gwa,Ding:2023ydy}.

It is also possible to build modular flavor models beyond the usual modular group $\SL{2,\mathds{Z}}$ by extending it to a generalized modular group.
As a first possibility, $\Gamma(N)$ has been considered instead of $\SL{2,\mathds{Z}}$ and some of the resulting finite symmetries obtained from quotients of $\Gamma(N)$ and its subgroups have been explored~\cite{Li:2021buv}. Further, one can extend $\SL{2,\mathds{Z}}$ to symplectic groups with Siegel modular forms~\cite{Ding:2020zxw} or metaplectic groups with half-integral-weight modular forms~\cite{Liu:2020msy}.
This enriches the possible constructions of modular flavor models that can be used for flavor phenomenology~\cite{Yao:2020zml,Ding:2021iqp,Ding:2023ydy,Arriaga-Osante:2023wnu,Ding:2024xhz,Ding:2024pix}.
These generalizations find additional motivation from their appearance in top-down constructions~\cite{Kobayashi:2018rad,Baur:2019kwi,Ishiguro:2020nuf,Kikuchi:2020nxn,Ohki:2020bpo,Baur:2020yjl,Nilles:2021glx,Almumin:2021fbk,Baur:2024qzo}.

It turns out that, in some cases, different quotients of the infinite modular groups lead to an identical finite modular groups.
For example, among the groups of order less than 72,
$T'$ and $\Z7\rtimes\Z6$ arise each from two different quotients of $\SL{2,\mathds{Z}}$ and its subgroups~\cite{Liu:2021gwa}.
Further, the finite modular group $T'$ arises from the quotient of $\Gamma(2)$ and $\Gamma(6)$ (and other quotients), and the \acp{VVMF} in this case are completely different from those obtained from quotients using $\SL{2,\mathds{Z}}$~\cite{Li:2021buv}.
This means that, even if the finite modular group is fixed, the modular symmetry is not.
In slightly more detail, for a given finite modular group $G$, the \acp{VVMF} can differ if different quotients are used to construct $G$.
This, in turn, leads to different Yukawa couplings.
It is the purpose of this paper to work out the differences, and explore their phenomenological consequences.

In this work, we start from the observation that the same finite modular group structure can be obtained from different quotients of modular groups.
These different quotients lead to different modular forms, and, as we shall see, different phenomenology.
As an example, we explore three realizations of the group $T'$ arising from the quotients
$\SL{2,\mathds{Z}}/\Gamma(3)$, $\SL{2,\mathds{Z}}/N_{[24,3]II}$ and $\Gamma(2)/\Gamma(6)$, where $N_{[24,3]II}$
is a normal subgroup of index 24 of $\SL{2,\mathds{Z}}$ different from $\Gamma(3)$.
\enlargethispage{\baselineskip}

We organize our paper as follows.
In \Cref{sec:VVMF} we briefly collect some basic facts on the framework of modular flavor symmetries.
In \Cref{sec:Tprime}, we provide details on three different scenarios that yield the modular flavor symmetry $T'$, including the different \acp{VVMF} arising in each case.
Finally, in \Cref{sec:models} we compare the quark phenomenology associated with these different $T'$ scenarios.
We conclude in \Cref{sec:conclusion} with a summary of our results and an outlook for future work.
Our appendices provide discussions on group theory for $T'$, \acp{VVMF} of the various realizations of modular $T'$, and details on the fits to data of our models.

\section{Modular symmetries and vector-valued modular forms}
\label{sec:VVMF}

In order to introduce the formalism and fix our notation, let us consider a particular supersymmetric nonlinear $\sigma$ model.
The chiral superfield content of the model is $\{\tau,\varphi^{(I)}\}$, where $\varphi^{(I)}$ denotes chiral matter field multiplets and $\tau$ refers here to the complex structure modulus of a $\mathds{T}^2$ torus.
$\tau$ takes values in the upper-half complex plane $\mathcal{H}$.\footnote{The upper-half plane is defined as $\mathcal{H}=\left\{\tau\in\mathds{C} ~|~\im\tau>0 \right\}$. It should be noted that we are often dealing with an extended upper half plane $\overline{\mathcal{H}}=\mathcal{H}\cup\mathds{Q}\cup\{\ii\infty\}$, which is the space expanded by adding some rational points and point at infinity to $\mathcal{H}$.}
The transformations that leave the torus $\mathds{T}^2$ invariant build the group $\Gamma = \SL{2,\mathds{Z}}$, generated by the two generators $S$ and $T$, which can be written as
\begin{equation}
	S=\begin{pmatrix}
		0 & 1 \\ -1 & 0
	\end{pmatrix}\;, \qquad
	T=\begin{pmatrix}
		1 & 1 \\ 0 & 1
	\end{pmatrix}\;.
\end{equation}
The modular transformations of $\varphi^{(I)}$ under $\Gamma$ involve a finite modular group given by $G\defeq\Gamma/G_d$, where $G_d$ is a normal subgroup of $\Gamma$ with finite index.
Thus, the supersymmetric nonlinear $\sigma$ model with modular symmetry, $\Gamma$ is defined by the data $(\Gamma, G_d\unlhd \Gamma)$~\cite{Liu:2021gwa}.
In particular, specifying only $G$ does, in general, \emph{not} fix the modular symmetry of a given model.
The modular transformations acting on the modulus and the chiral matter fields are given by
\begin{subequations}\label{eq:modular_transform}
\begin{align}
		\tau&\xmapsto{~\gamma~}  \dfrac{a \tau+b}{c \tau+d}\;,\\
		\varphi^{(I)}&\xmapsto{~\gamma~}(c\tau+d)^{-k_I}\rho^{(I)}(\gamma)\varphi^{(I)}\;,
\end{align} 
with
\begin{equation}
	\gamma=\begin{pmatrix}
		a  &  b  \\
		c  &  d
	\end{pmatrix}\in \Gamma\
	\quad\text{and}\quad\rho\in\operatorname{Rep}(\Gamma/G_d)\;.
\end{equation}
\end{subequations}
Here, $a,b,c,d\in\mathds{Z}$ are subject to $a\,d-b\,c=1$, and $(c\tau+d)^{-k_I}$ is known as the automorphy factor of modular weight $k_I$ associated with the field multiplet $\varphi^{(I)}$.
Note that the finite modular group $\Gamma/G_d$ can also be inferred from the modular transformations of $\varphi^{(I)}$ since it can also be defined as $\image(\rho)\cong \Gamma/\ker(\rho)$ in terms of the representation $\rho$.
That is, as $\ker(\rho)$ coincides with the normal subgroup $G_d \unlhd \Gamma$ with finite index, specifying $\rho$ amounts to setting $G_d$.
Thus, the \ac{SUSY} nonlinear $\sigma$ model with modular symmetry $\Gamma$ can also be defined by the data $(\Gamma,\rho)$.

$\mathrm{SL}(2,\mathds{Z})$ has many normal subgroups $G_d$.
A common class of subgroups are the so-called principal congruence subgroups of level $n$, defined as
\begin{equation}
	\label{eq:principalCongruence}
	\Gamma(n)=\left\{\gamma \in \Gamma ~|~ \gamma \equiv\mathds{1}\bmod n \,\right\}\;.
\end{equation}
The resulting quotients build the homogeneous finite modular groups $\Gamma'_n\defeq\Gamma/\Gamma(n)$, which are isomorphic to the discrete groups $S_3$, $T'$, $S_4'$, and $A_5'$ for $n=2$ , $3$, $4$, and $5$, respectively.
However, there are many $G_d$ that do not fall into this category~\cite{Liu:2021gwa}.
For example, it is found that there are two normal subgroups of finite index 24 that result in $\Gamma/G_d \cong [24,3]$ in the conventions of GAP~\cite{GAP}, where the first number in the brackets is the order of the finite group and the second number is an additional label.
In our notation, the two normal subgroups are $N_{[24,3]I} = \Gamma(3)$ and $N_{[24,3]II}$, where the latter does not belong to the set of principal congruence subgroups.
The GAP Id $[24,3]$ corresponds to the group traditionally called $T'\cong\SL{2,\mathds{Z}_3}$.

In the framework of modular flavor symmetries, the Yukawa couplings between the matter fields in supersymmetric models get promoted to \acp{VVMF}~\cite{Feruglio:2017spp}.
Since modular forms of $\Gamma$ are uniquely determined, it is possible to work out the properties of the \acp{VVMF} for all representations $\rho$.
Under the action of an arbitrary modular transformation $\gamma\in\Gamma$, the $r$-dimensional \acp{VVMF} $Y^{(k_Y)}_{\rep{r}}(\tau)$ of modular weight $k_Y$ transform as
\begin{equation}
	Y^{(k_Y)}_{\rep{r}}(\tau) \xmapsto{~\gamma~} Y^{(k_Y)}_{\rep{r}}(\gamma\tau) = (c\tau+d)^{k_Y}\rho^{(Y)}(\gamma)Y^{(k_Y)}_{\rep{r}}(\tau)\;,
\end{equation}
where $\rho^{(Y)}(\gamma)$ denotes the corresponding representation of $\gamma$ in the finite modular group $\Gamma/G_d$.

In the bottom-up approach, one usually assumes global \ac{SUSY} along with a superpotential $\mathscr{W}(\tau,\varphi^{(I)})$,
which is required to be invariant under the modular transformations~\eqref{eq:modular_transform}.
This implies that any admissible superpotential term is subject to a condition on the modular weights and the representations of matter multiplets appearing in the couplings, i.e.\
\begin{equation}
  \mathscr{W}(\tau,\varphi^{(I)}) \supset Y^{(k_Y)}_{\rep{r}}(\tau) \prod_I \varphi^{(I)}
  \qquad\iff\qquad
  k_Y \overset{!}{=} \sum_I k_I\quad \text{and}\quad
  \rep{1} \overset{!}{\in} \rho^{(Y)}\bigotimes _I \rho^{(I)} \;,
\end{equation}
where $\rep1$ denotes the trivial singlet of $\Gamma/G_d$.

The present discussion can be extended to more general supersymmetric nonlinear $\sigma$ models with similar matter content $\{\tau,\varphi^{(I)}\}$, where the modular transformations arise from a generalized modular group $\Upsilon$.
As long as the moduli space is a subset of $\overline{\mathcal{H}}$,\footnote{More precisely, the moduli space here refers to the fundamental domain of a certain infinite discrete group $\Upsilon$, $\overline{\mathcal{H}}/\Upsilon$, that is, the set of inequivalent points in the extended upper half plane $\overline{\mathcal{H}}$ that cannot be related to each other by transformations under the group $\Upsilon$.}
the model is specified by the data $( \Upsilon, G_d\unlhd \Upsilon)$.
We consider $\Upsilon$ to be some infinite discrete modular group acting on $\tau\in\mathcal{H}$, which includes but is not limited to the principal congruence subgroups~\eqref{eq:principalCongruence}.
In this case, the modular transformations of the modulus and matter fields are still given by \Cref{eq:modular_transform}
with $\gamma\in\Upsilon$ and $\rho$ a representation of the quotient group $\Upsilon/G_d$.
Interestingly, if one chooses $\Upsilon=\Gamma(2)$ and considers $G_d=\Gamma(6)\unlhd\Gamma(2)$, the resulting finite modular symmetry is $\Upsilon/G_d = \Gamma(2)/\Gamma(6)\cong[24,3]$ as well~\cite{Li:2021buv}.
This reveals in particular that the origin of the finite modular group $T'$ is far from unique.

\section{Various instances of \texorpdfstring{$T'$}{T(prime)} and their modular forms}
\label{sec:Tprime}

\subsection{\texorpdfstring{$T'$}{T(prime)} as a generalized finite modular group of \texorpdfstring{$\SL{2,\mathds{Z}}$}{SL(2,Z)}}
\label{sec:TprimeFromGamma}

As mentioned earlier, in the sequence of generalized finite modular groups over $\SL{2,\mathds{Z}}$, there exist two finite modular groups isomorphic to $T'$.
One of them is the homogeneous finite modular group $\Gamma'_3 = \Gamma / \Gamma(3)$, whose presentation is given by
\begin{equation}
\label{eq:GammaPrime3Presentation}
   \Gamma'_3 = \Braket{~S,T~|~S^4=(S\,T)^3=T^3=\Id,~S^2T=T\,S^2~} \cong T'\;.
\end{equation}
The second instance of $T'$ arises from the quotient of $\SL{2,\mathds{Z}}$ divided by the (infinite modular) normal subgroup,
\begin{equation}
\label{eq:OtherNormalSubgroup}
    N_{[24,3]II}\defeq\Braket{~S^{2}\,T^{3},~S\,T^{3}S,~S\,T\,S\,T^{-2}\,S^{-1}\,T~}\;.
\end{equation}
Here, the elements in the brackets are the generators.
In order to distinguish the resulting finite modular group from $\Gamma'_3$, we denote it as
\begin{equation}
\label{eq:AnotherTPrimePresentation}
   2T\defeq \Gamma/N_{[24,3]II} = \Braket{~S,T~|~S^4=(S\,T)^3=S^2\,T^3=\Id,~S^2\,T=T\,S^2~} \cong T'\;.
\end{equation}
Note that the \acp{irrep} of $\Gamma'_3$ and $2T$, as presented in \Cref{tab:Tprime_reps}, differ.

\begin{table}[t!]
 \centering
 \begin{tabular}{ccc}
	\toprule
 $\rep{r}$   & $\rho_{\Gamma'_3}(S)$ & $\rho_{\Gamma'_3}(T)$ \\ 
 \midrule
 $\rep{1}$   & $1$                   & $1$                   \\ 
 $\rep{1}'$  & $1$                   & $\omega$              \\ 
 $\rep{1}''$ & $1$                   & $\omega^2$            \\ 
 \cellcolor{lightgray}
 $\rep{2}$   & \cellcolor{lightgray}
 $\frac{\ii}{\sqrt3}
            \begin{pmatrix}
                1      & \sqrt2  \\
                \sqrt2 & -1
            \end{pmatrix}$ & \cellcolor{lightgray}
            $\omega
            \begin{pmatrix}
                1 & 0 \\
                0 & \omega
            \end{pmatrix}$ \\ 
\cellcolor{lightgray}$\rep{2}'$  & \cellcolor{lightgray}$\frac{\ii}{\sqrt3}
            \begin{pmatrix}
                1      & \sqrt2 \\
                \sqrt2 & -1
            \end{pmatrix}$ & \cellcolor{lightgray}
            $\omega^2
            \begin{pmatrix}
                1 & 0 \\
                0 & \omega
            \end{pmatrix}$ \\ 
\cellcolor{lightgray} $\rep{2}''$ & \cellcolor{lightgray} $\frac{\ii}{\sqrt3}
            \begin{pmatrix}
                1      & \sqrt2 \\
                \sqrt2 & -1
            \end{pmatrix}$ & \cellcolor{lightgray}
            $
            \begin{pmatrix}
                1 & 0 \\
                0 & \omega
            \end{pmatrix}$ \\ 
$\rep{3}$  & $\frac{1}{3}
            \begin{pmatrix}
                -1 & 2 & 2 \\
                2 & -1 & 2 \\
                2 & 2 & -1
            \end{pmatrix}$ &
            $
            \begin{pmatrix}
                1 & 0 & 0\\
                0 & \omega & 0\\
                0 & 0 & \omega^2
            \end{pmatrix}$ \\
            \bottomrule
\end{tabular}
\hspace{1cm}%
\begin{tabular}{ccc}
	\toprule
 $\rep{r}$   & $\rho_{2T}(S)$ & $\rho_{2T}(T)$ \\ 
 \midrule
 $\rep{1}$   & $1$            & $1$            \\ 
 $\rep{1}'$  & $1$            & $\omega$       \\ 
 $\rep{1}''$ & $1$            & $\omega^2$     \\ 
 \cellcolor{lightgray}$\rep{2}$   & \cellcolor{lightgray}$\frac{-\ii}{\sqrt3}
            \begin{pmatrix}
                1      & \sqrt2\\
                \sqrt2 & -1
            \end{pmatrix}$ & \cellcolor{lightgray}
            $-\omega
            \begin{pmatrix}
                1 & 0 \\
                0 & \omega
            \end{pmatrix}$ \\ 
\cellcolor{lightgray}$\rep{2}'$ & \cellcolor{lightgray}$\frac{-\ii}{\sqrt3}
            \begin{pmatrix}
                1      & \sqrt2 \\
                \sqrt2 & -1
            \end{pmatrix}$ & \cellcolor{lightgray}
            $-\omega^2
            \begin{pmatrix}
                1 & 0 \\
                0 & \omega
            \end{pmatrix}$ \\ 
\cellcolor{lightgray}$\rep{2}''$ & \cellcolor{lightgray}$\frac{-\ii}{\sqrt3}
            \begin{pmatrix}
                1 & \sqrt2 \\
                \sqrt2 & -1
            \end{pmatrix}$ & \cellcolor{lightgray}
            $-
            \begin{pmatrix}
                1 & 0 \\
                0 & \omega
            \end{pmatrix}$ \\ 
$\rep{3}$  & $\frac{1}{3}
            \begin{pmatrix}
                -1 & 2 & 2 \\
                2 & -1 & 2 \\
                2 & 2 & -1
            \end{pmatrix}$ &
            $
            \begin{pmatrix}
                1 & 0 & 0\\
                0 & \omega & 0\\
                0 & 0 & \omega^2
            \end{pmatrix}$ \\
            \bottomrule
\end{tabular}
\caption{\label{tab:Tprime_reps}%
Representation matrices of the generators $S$ and $T$ for the finite modular group $\Gamma'_3\cong T'$ and $2T\cong T'$, where $\omega=\ee^{2\pi \ii / 3}$.
Note that all 1- and 3-dimensional representation matrices of $\Gamma'_3$ and $2T$ are identical, while the 2-dimensional representations (highlighted in gray) differ by a minus sign.
The \ac{CG} coefficients corresponding to these two instances of $T'$ are the same, and they are given in \Cref{app:GroupTprime}.}
\end{table}

Since $\Gamma_3'$ and $2T$ are both isomorphic to $T'$, i.e.\ since they have the same group structure, we can redefine the generators of $2T$ to bring the presentation~\eqref{eq:AnotherTPrimePresentation} to the standard modular $T'$ presentation~\eqref{eq:GammaPrime3Presentation}.
We find that
\begin{equation}
\label{eq:TprimeRelations}
   2T = \Braket{~s\defeq S^3,t\defeq T^2S~|~s^4=(s\,t)^3=t^3=\Id,~s^2t=t\,s^2~} \cong T'\;.
\end{equation}
Note that the $2T$ representation matrices $\rho_{2T}(s)=\rho_{2T}(S)^3$ and $\rho_{2T}(t)=\rho_{2T}(T)^2\rho_{2T}(S)$ satisfy the relators given in \Cref{eq:TprimeRelations}.
Despite the equivalence of the presentations, the $\Gamma_3'$ generators $S$ and $T$ and the $2T$ generators $s$ and $t$ act differently on the modulus and matter fields,
\begin{subequations}
	\begin{align}
		\Gamma'_3:&~~
		\left\{\begin{aligned}
			\tau     &   \xmapsto{~S~} -1/\tau\;, &
			\tau     &  \xmapsto{~T~} \tau+1\;, \\
			\varphi^{(I)} & \xmapsto{~S~} (-\tau)^{-k_I}\rho^{(I)}_{\Gamma'_3}(S)\,\varphi^{(I)}\;,\qquad&
			\varphi^{(I)} & \xmapsto{~T~} \rho^{(I)}_{\Gamma'_3}(T)\,\varphi^{(I)}\;,
		\end{aligned}
		\right.
		\\
		2T:&~~ \left\{\begin{aligned}
			\tau &          \xmapsto{~s=S^3~} -1/\tau\;,
			&\tau         & \xmapsto{~t=T^2S~} 2-1/\tau \;,\\
			\varphi^{(I)} & \xmapsto{~s=S^3~} (\tau)^{-k_I}\rho^{(I)}_{2T}(s)\,\varphi^{(I)}\;,\quad&
			\varphi^{(I)} &\xmapsto{~t=T^2S~} (-\tau)^{-k_I}\rho^{(I)}_{2T}(t)\,\varphi^{(I)}\;.
		\end{aligned}
		\right.
	\end{align}
\end{subequations}
On the other hand, if we do not redefine the $2T$ generators, then the two groups act differently only on matter fields building doublets because of the different representation matrices of \Cref{tab:Tprime_reps}.
We see that (we repeat the $\Gamma'_3$ transformations for the sake of comparison)
\begin{subequations}
	\begin{align}
		\Gamma'_3:&~~\left\{\begin{aligned}
			\tau         &  \xmapsto{~S~} -1/\tau\;, \quad
			&\tau        &  \xmapsto{~T~} \tau+1\;, \\
			\varphi^{(I)} & \xmapsto{~S~} (-\tau)^{-k_I}\rho^{(I)}_{\Gamma'_3}(S)\,\varphi^{(I)}\;,&\qquad
			\varphi^{(I)} &\xmapsto{~T~} \rho^{(I)}_{\Gamma'_3}(T)\,\varphi^{(I)}\;,
		\end{aligned}
		\right.\\
		2T:&~~ \left\{\begin{aligned}
			\tau           &\xmapsto{~S~} -1/\tau\;, \quad
			&\tau          &\xmapsto{~T~} \tau+1\;, \\
			\varphi^{(I)}  &\xmapsto{~S~} (-\tau)^{-k_I} \rho^{(I)}_{2T}(S)\,\varphi^{(I)}\;,\qquad
			&\varphi^{(I)}& \xmapsto{~T~} \rho^{(I)}_{2T}(T)\,\varphi^{(I)}\;.
		\end{aligned}
		\right.
	\end{align}
\end{subequations}

\begin{table}[t!]
	{\centering
		\begin{tabular}{c@{\;\;}c@{\;\;}c@{\;\;}c}
			\toprule
			$\rep{r}$   & $k_0$ & Analytical expression & $q$-expansion \\ 
			\midrule
			$\rep{1}$   & 0     & $1$ & $1$ \\ 
			$\rep{1}'$  & 4     & $\eta^{8}$ & $q^{1/3}(1-8 q+20 q^2-70 q^4+ \cdots)$ \\ 
			$\rep{1}''$ & 8     & $\eta^{16}$ & $q^{2/3}(1-16 q+104 q^2-320 q^3+260 q^4+ \cdots)$ \\ 
			$\cellcolor{lightgray}\rep{2}$ &\cellcolor{lightgray} 5 &
			\cellcolor{lightgray}
			$\begin{pmatrix}
				\eta^{10}\left(\frac{K}{1728}\right)^{-\frac{1}{12}}{}_2F_1\left(-\frac{1}{12},\frac{1}{4};\frac{2}{3};K\right) \\
				3\sqrt{2}\eta^{10}\left(\frac{K}{1728}\right)^{\frac{1}{4}}{}_2F_1\left(\frac{1}{4},\frac{7}{12};\frac{4}{3};K\right) \\
			\end{pmatrix}$
			&
			\cellcolor{lightgray}
			$\begin{pmatrix}
				q^{1/3}(1-2 q-28 q^2+126 q^3-112 q^4+\cdots) \\
				3\sqrt{2}q^{2/3}(1-7 q+14 q^2+4 q^3-28 q^4+\cdots)\\
			\end{pmatrix}$
			\\ 
			$\cellcolor{lightgray}\rep{2}'$ & \cellcolor{lightgray}3 &
			\cellcolor{lightgray}
			$\begin{pmatrix}
				\eta^{6}\left(\frac{K}{1728}\right)^{\frac{5}{12}}{}_2F_1\left(\frac{5}{12},\frac{3}{4};\frac{5}{3};K\right) \\
				\frac{-\sqrt{2}\,\eta^{6}}{54}\left(\frac{K}{1728}\right)^{-\frac{1}{4}}{}_2F_1\left(-\frac{1}{4},\frac{1}{12};\frac{1}{3};K\right) \\
			\end{pmatrix}$
			&
			\cellcolor{lightgray}
			$\begin{pmatrix}
				q^{2/3}(1+8 q+17 q^2+40 q^3+50 q^4+\cdots) \\
				\frac{-\sqrt{2}}{54}(1+72q+270 q^2+720 q^3+936 q^4
				+\cdots)\\
			\end{pmatrix}$
			\\ 
			\cellcolor{lightgray} $\rep{2}''$ & \cellcolor{lightgray}1 &
			\cellcolor{lightgray}
			$\begin{pmatrix}
				\eta^{2}\left(\frac{K}{1728}\right)^{-\frac{1}{12}}{}_2F_1\left(-\frac{1}{12},\frac{1}{4};\frac{2}{3};K\right) \\
				3\sqrt{2}\eta^{2}\left(\frac{K}{1728}\right)^{\frac{1}{4}}{}_2F_1\left(\frac{1}{4},\frac{7}{12};\frac{4}{3};K\right) \\
			\end{pmatrix}$
			&
			\cellcolor{lightgray}
			$\begin{pmatrix}
				1+6 q+6 q^3+6 q^4+\cdots \\
				3\sqrt{2}q^{1/3}(1+q+2q^2+2 q^4+\cdots)\\
			\end{pmatrix}$
			\\ 
			$\rep{3}$ & 2 &
			$\begin{pmatrix}
				\eta^{4}\left(\frac{K}{1728}\right)^{-\frac{1}{6}}{}_3F_2\left(-\frac{1}{6},\frac{1}{6},\frac{1}{2};\frac{2}{3},\frac{1}{3};K\right) \\
				-6\eta^{4}\left(\frac{K}{1728}\right)^{\frac{1}{6}}{}_3F_2\left(\frac{1}{6},\frac{1}{2},\frac{5}{6};\frac{2}{3},\frac{4}{3};K\right) \\
				-18 \eta^{4}\left(\frac{K}{1728}\right)^{\frac{1}{2}}{}_3F_2\left(\frac{1}{2},\frac{5}{6},\frac{7}{6};\frac{5}{3},\frac{4}{3};K\right) \\
			\end{pmatrix}$
			&
			$\begin{pmatrix}
				1+12 q+36 q^2+12 q^3+84 q^4
				+\cdots \\
				-6 q^{1/3}(1+7 q+8 q^2+18 q^3+14 q^4+\cdots)\\
				-18 q^{2/3}(1+2 q+5 q^2+4 q^3+8 q^4 + \cdots)
			\end{pmatrix}$
			\\ \bottomrule
		\end{tabular}
		\caption{\label{tab:GammaPrime3_modularForm}%
		Basis of \acp{VVMF} of the finite modular group $\Gamma'_3\cong[24,3]$ with minimal weight $k_0$ for the various representations of the finite modular group.
		$\eta$ denotes the Dedekind eta function, $K(\tau)\defeq1728/j(\tau)$ in terms of the modular $j$-invariant~\cite{cohen2017modular}, $\prescript{}{2}F_1$ and $\prescript{}{3}F_2$ are hypergeometric functions, and $q\defeq\ee^{2\pi\ii\tau}$.
		We highlight the modular forms building doublets with a light-gray background because they differ from those associated with $2T$, cf.~\Cref{tab:2T_modularForm}.}
	}
\end{table}
\begin{table}[h!]
	{\centering
		\begin{tabular}{c@{\;\;}c@{\;\;}c@{\;\;}c}
			\toprule
			$\rep{r}$ & $k_0$ & Analytical expression & $q$-expansion \\ 
			\midrule
			$\rep{1}$ & 0 & $1$ & $1$ \\ 
			$\rep{1}'$ & 4 & $\eta^{8}$ & $q^{1/3}(1-8 q+20 q^2-70 q^4+ \cdots)$ \\ 
			$\rep{1}''$ & 8 & $\eta^{16}$ & $q^{2/3}(1-16 q+104 q^2-320 q^3+260 q^4+ \cdots)$ \\ 
			\cellcolor{lightgray}$\rep{2}$ &\cellcolor{lightgray} 5 &
			\cellcolor{lightgray}
			$\begin{pmatrix}
				\eta^{10}\left(\frac{K}{1728}\right)^{\frac{5}{12}}{}_2F_1\left(\frac{5}{12},\frac{3}{4};\frac{5}{3};K\right) \\
				\frac{-\eta^{10}}{27\sqrt{2}}\left(\frac{K}{1728}\right)^{-\frac{1}{4}}{}_2F_1\left(-\frac{1}{4},\frac{1}{12};\frac{1}{3};K\right) \\
			\end{pmatrix}$
			&
			\cellcolor{lightgray}
			$\begin{pmatrix}
				q^{5/6}(1+4 q-13 q^2-4 q^3-17 q^4+\cdots)\\
				\frac{-q^{1/6}}{27\sqrt{2}}(1+68 q-16 q^2-208 q^3-833 q^4+\cdots) \\
			\end{pmatrix}$
			\\ 
			\cellcolor{lightgray}$\rep{2}'$ & \cellcolor{lightgray}3 &
			\cellcolor{lightgray}
			$\begin{pmatrix}
				\eta^{6}\left(\frac{K}{1728}\right)^{-\frac{1}{12}}{}_2F_1\left(-\frac{1}{12},\frac{1}{4};\frac{2}{3};K\right)  \\
				3\sqrt{2}\eta^{6}\left(\frac{K}{1728}\right)^{\frac{1}{4}}{}_2F_1\left(\frac{1}{4},\frac{7}{12};\frac{4}{3};K\right)\\
			\end{pmatrix}$
			&
			\cellcolor{lightgray}
			$\begin{pmatrix}
				q^{1/6}(1+2 q-22 q^2+26q^3+25 q^4+\cdots)\\
				3\sqrt{2}q^{1/2}(1-3 q+2 q^3+9 q^4+\cdots) \\
			\end{pmatrix}$
			\\ 
			$\cellcolor{lightgray}\rep{2}''$ &\cellcolor{lightgray}7 &
			\cellcolor{lightgray}
			$\begin{pmatrix}
				\eta^{14}\left(\frac{K}{1728}\right)^{-\frac{1}{12}}{}_2F_1\left(-\frac{1}{12},\frac{1}{4};\frac{2}{3};K\right) \\
				3\sqrt{2}\eta^{14}\left(\frac{K}{1728}\right)^{\frac{1}{4}}{}_2F_1\left(\frac{1}{4},\frac{7}{12};\frac{4}{3};K\right)\\
			\end{pmatrix}$
			&
			\cellcolor{lightgray}
			$\begin{pmatrix}
				q^{1/2}(1-6 q-18 q^2+242 q^3-693 q^4+\cdots)\\
				3\sqrt{2}q^{5/6}(1-11 q+44 q^2-58 q^3-77 q^4+\cdots)\\
			\end{pmatrix}$
			\\ 
			$\rep{3}$ & 2 &
			$\begin{pmatrix}
				\eta^{4}\left(\frac{K}{1728}\right)^{-\frac{1}{6}}{}_3F_2\left(-\frac{1}{6},\frac{1}{6},\frac{1}{2};\frac{2}{3},\frac{1}{3};K\right) \\
				-6\eta^{4}\left(\frac{K}{1728}\right)^{\frac{1}{6}}{}_3F_2\left(\frac{1}{6},\frac{1}{2},\frac{5}{6};\frac{2}{3},\frac{4}{3};K\right) \\
				-18\eta^{4}\left(\frac{K}{1728}\right)^{\frac{1}{2}}{}_3F_2\left(\frac{1}{2},\frac{5}{6},\frac{7}{6};\frac{5}{3},\frac{4}{3};K\right) \\
			\end{pmatrix}$
			&
			$\begin{pmatrix}
				1+12 q+36 q^2+12 q^3+84 q^4
				+\cdots \\
				-6q^{1/3}(1+7 q+8 q^2+18 q^3+14 q^4+\cdots)\\
				-18q^{2/3}(1+2 q+5 q^2+4 q^3+8 q^4 + \cdots)
			\end{pmatrix}$
			\\ \bottomrule
		\end{tabular}%
		\caption{\label{tab:2T_modularForm}Basis of \acp{VVMF} of the finite modular group $2T$ with minimal weight $k_0$ for all possible \acp{irrep}. \acp{VVMF} building doublets are highlighted in light gray, as they differ from the ones for $\Gamma_3'$, see \Cref{tab:GammaPrime3_modularForm}.}
	}
\end{table}

Given this information, the \acp{VVMF} transforming under the representations of $\Gamma_3'$ and $2T$ can be obtained by the method of the \ac{MLDE}~\cite{Liu:2021gwa, bantay2007vector}.
This implies, in particular, that the \acp{VVMF} associated with $\Gamma_3'$ and with $2T$ that transform as doublets of those groups are different. 
Following the methods described in~\cite{Liu:2021gwa}, we compute the \acp{VVMF} of minimal weight $k_0$ for each of the possible \acp{irrep} of a given finite group.
The resulting \acp{VVMF} are given in \Cref{tab:GammaPrime3_modularForm} for $\Gamma_3'$ and in \Cref{tab:2T_modularForm} for $2T$, where we provide both their analytical expressions and their $q$-expansions in terms of $q\defeq\ee^{2\pi\ii\tau}$.

The structure of \acp{VVMF} allows one to build higher-weight modular forms either (i) by multiplying polynomials of $E_4$ and $E_6$ and applying the modular differential operator $D^n_{k_0}$ into these base modular forms,\footnote{The modular differential operator $D^n_{k_0}$ is defined by $D^{n}_{k_0} \defeq D_{{k_0}+2(n-1)}\circ D_{{k_0}+2(n-2)}\circ\dots \circ D_{k_0}\,$, where the modular derivative $D_{k_0}$ is defined by  $D_{k_0} \defeq \frac{1}{2\pi\ii} \frac{\dd}{\dd\tau}- \frac{{k_0}E_2(\tau)}{12}$.} along the lines described in~\cite{Liu:2021gwa}, or (ii) by tensor products of the elements of the basis of \acp{VVMF} listed in our tables.
For example, the weight-3 doublet $Y^{(3)}_{\rep{2''}}$ of $\Gamma'_{3}$ is given by
\begin{equation}
\label{eq:weight3doublet2''}
   Y^{(3)}_{\rep{2}''} = \left(Y^{(1)}_{\rep{2''}} Y^{(2)}_{\rep{3}}\right)_{\rep{2}''} = -12 D_1 Y^{(1)}_{\rep{2''}}\;,
\end{equation}
and the corresponding $q$-expansion is
\begin{equation}
\label{eq:weight3doublet2''_qExp}
    Y^{(3)}_{\rep{2}''}=
    \begin{pmatrix} 1 - 90 q - 216 q^2 - 738 q^3 - 1170 q^4 + \cdots \\
                   -9 \sqrt{2} q^{1/3}\left(1 + 13 q + 50 q^2 + 72 q^3+ \cdots \right)
    \end{pmatrix}\;.
\end{equation}

\subsection{\texorpdfstring{$T'$}{T(prime)} as a finite modular group of \texorpdfstring{$\Gamma(2)$}{Gamma(2)}}
\label{sec:TprimeFromGamma2}

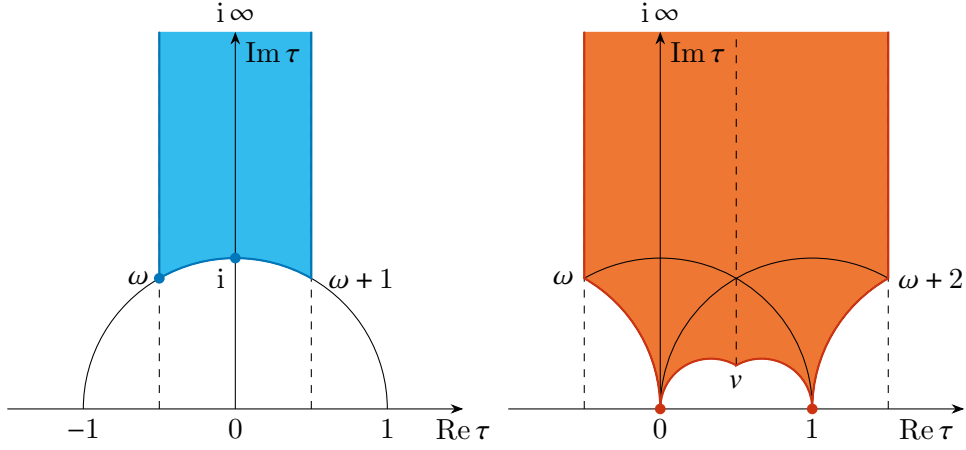
\begin{figure}[t!]
	\centering
	\begin{tikzpicture}[>=Stealth,declare function={R=2;}]

		\draw [->] (-3,0) -- (3,0) node [below] {$\operatorname{Re}\tau$};

		\draw[->] (0,0) -- (0,5) node[below right]{$\operatorname{Im}\tau$} node[above] {$\ii\infty$};

		\path (0,5) coordinate (T);

		\path (0,0) coordinate (0) node[below] {$0$} ;
		\draw[radius=R] (R,0) coordinate (1) node[below]{$1$}
		arc[start angle=0,end angle=60] coordinate (omega2) node[right]{$~\omega+1$} ;

		\draw[radius=R,draw=c1] (omega2) arc[start angle=60,end angle=90] coordinate (i) node[below left]{$\ii$} arc[start angle=90,end angle=120] coordinate (omega) node[left]{$\omega$};

		\draw[radius=R] (omega) arc[start angle=120,end angle=180] coordinate (-1) node[below] {$-1$};

		\draw[dashed] (omega|-0) -- (omega|-T) (omega2|-0) -- (omega2|-T);

		\begin{pgfonlayer}{background layer}

			\filldraw[fill=c2] (1,5) -- (omega2) arc (60:120:R) -- (omega) -- (-1,5) ;

		\end{pgfonlayer}

		\draw[thick,draw=c1] (1,5) -- (omega2) arc (60:120:R) -- (omega) -- (-1,5) ;

		\fill [c1] foreach \p in {omega,i} { (\p) circle[radius=2pt]};

	\end{tikzpicture}
	\begin{tikzpicture}[>=Stealth,declare function={R=2;r=2/3;}]

		\draw[->] (-2,0) -- (4,0) node[below left]{$\operatorname{Re}\tau$};

		\draw[->] (0,0) -- (0,5) node[below right]{$\operatorname{Im}\tau$}  node[above] {$\ii\infty$};

		\path (0,5) coordinate (T);

		\path (0,4.7) coordinate (INF) ;

		\draw[radius=R] (R,0) coordinate (1) node[below]{$1$}
		arc[start angle=0,end angle=120] coordinate (omega) node[left]{$\omega$}
		(1) arc[start angle=180,end angle=120,color=c5]
		(0,0) coordinate (0) node[below]{$0$}
		arc[start angle=180,end angle=60] coordinate (omega2) node[right]{$\omega+2$}
		(0) arc[start angle=0,end angle=60]
		;

		\draw[radius=R,c5] (omega) arc[start angle=60,end angle=0] (0) ;

		\draw[radius=R,c5] (1) arc[start angle=180,end angle=120] (omega2) ;

		\draw[radius=r,c5] (0) arc[start angle=180,end angle=60] coordinate (v) node[below,black] {$v$} arc[start angle=120,end angle=0];

		\draw[dashed] (omega|-0) -- (omega|-T) (omega2|-0) -- (omega2|-T) (v) -- (v|-T);

		\draw[-,c5] (omega) -- (omega|-T) (omega2) -- (omega2|-T) ;

		\begin{pgfonlayer}{background layer}

			\filldraw[fill=c4] (-1,5) -- (omega) arc (60:0:R) -- (0) arc (180:60:r) -- (v) arc (120:0:r) -- (1) arc (180:120:R) -- (omega2) -- (3,5) ;

		\end{pgfonlayer}

		\draw[thick,draw=c5] (-1,5) -- (omega) arc (60:0:R) -- (0) arc (180:60:r) -- (v) arc (120:0:r) -- (1) arc (180:120:R) -- (omega2) -- (3,5) ;

		\fill [c5] foreach \p in {0,1} { (\p) circle[radius=2pt]};

	\end{tikzpicture}
	\caption{\label{fig:fundamentaldomain}
	Fundamental domains of \SL{2,\mathds{Z}} (left) and $\Gamma(2)$ (right), including their critical points as colored dots.
	Here, $v\defeq(1-\omega)^{-1}=\frac{1}{2}+\ii\frac{\sqrt{3}}{6}$.}
\end{figure}

The group $T'\cong[24,3]$ can also be obtained as the quotient $\Gamma(2)/\Gamma(6)\cong T'$, as shown in \cite{Li:2021buv}.
In contrast to the traditional $T'$ presentation, \Cref{eq:GammaPrime3Presentation}, in this case the $S$ transformation
cannot be a generator of the finite modular group as $S$ does not appear in $\Gamma(2)$.
In this case, the abstract presentation of the finite modular group is given by
\begin{equation}
   \label{eq:Gamma2overGamma6Presentation}
   \Gamma(2) / \Gamma(6) = \Braket{~a,b~|~a^4=(a\,b)^3=b^3=\Id\;,a^2b=ba^2~} \cong T'\;,
\end{equation}
where $a\defeq T\,S\,T^4\,S^3T$ and $b\defeq T^4$ are the generators of $\Gamma(2)$ and $S$, $T$ are the generators of $\SL{2,\mathds{Z}}$.
The corresponding \acp{irrep} for this realization of $T'$ are shown in \Cref{tab:Quotient_reps}.
Another relevant difference is that the modulus $\tau$ must lie within the fundamental domain of $\Gamma(2)$, as shown in~\Cref{fig:fundamentaldomain}.

A way to determine the \acp{VVMF} corresponding to $\operatorname{Rep}\bigl(\Gamma(2)/\Gamma(6)\bigr)$ consists in first computing the scalar modular forms of $\operatorname{Rep}\bigl(\Gamma/\Gamma(6)\bigr)$ and then finding their transformation properties under $\Gamma(2)$.
Following this procedure, it is found that the modular forms of weight 1 of $\operatorname{Rep}\bigl(\Gamma/\Gamma(6)\bigr)$ are given by~\cite{Li:2021buv}
\begin{subequations}
	\begin{align}
		e_1(\tau) &\defeq \frac{\eta^3(3\tau)}{\eta(\tau)}\;, &
		e_2(\tau) &\defeq \frac{\eta^3(\tau/3)}{\eta(\tau)}\;, &
		e_3(\tau) &\defeq \frac{\eta^3(6\tau)}{\eta(2\tau)}\;, \\
		e_4(\tau) &\defeq \frac{\eta^3(\tau/6)}{\eta(\tau/2)}\;, &
		e_5(\tau) &\defeq \frac{\eta^3(2\tau/3)}{\eta(2\tau)}\;, &
		e_6(\tau) &\defeq \frac{\eta^3(3\tau/2)}{\eta(\tau/2)}\;.
	\end{align}
\end{subequations}
These are arranged as the three doublets of the group $T'\cong[24,3]$ according to
\begin{subequations}
\label{eq:weight1FormsG6}
	\begin{align}
		Y_{\rep{2}''}^{(1)}(\tau)&\defeq\begin{pmatrix}Y_{1}\\ Y_{2}\end{pmatrix} = \begin{pmatrix}3e_{1}(\tau)+e_{2}(\tau)\\ 3\sqrt{2}e_{1}(\tau)\end{pmatrix}\;,\\
		Y_{\rep{2}'I}^{(1)}(\tau)&\defeq\begin{pmatrix}Y_{3}\\ Y_{4}\end{pmatrix} = \begin{pmatrix}3\sqrt{2}e_{3}(\tau)\\ -3e_{3}(\tau) - e_{5}(\tau)\end{pmatrix}\;,\\
		Y_{\rep{2}'II}^{(1)}(\tau)&\defeq\begin{pmatrix}Y_{5}\\ Y_{6}\end{pmatrix}= \begin{pmatrix}\sqrt{6}\left(e_{3}(\tau)-e_{6}(\tau)\right) \\ 
			-\sqrt{3}e_{3}(\tau)+\frac{1}{\sqrt{3}}e_{4}(\tau)-\frac{1}{\sqrt{3}}e_{5}(\tau)+\sqrt{3}e_{6}(\tau)\end{pmatrix}\;,
	\end{align}
\end{subequations}
whose $q$-expansions are given by
\begin{subequations}
\label{eq:weight1FormsG6qexpansions}
	\begin{align}
	Y_{\rep{2''}}^{(1)}&=\begin{pmatrix}~1 + 6 q + 6 q^3 + 6 q^4 + 12 q^7 + 6 q^9 + \cdots~\\
	~3 \sqrt{2}q^{1/3} (1 + q + 2 q^2 + 2 q^4 + q^5 + 2 q^6 + q^8 + 2 q^9 + 2 q^{10} + \cdots) ~\end{pmatrix}\;,\\
	Y_{\rep{2'}I}^{(1)}&=\begin{pmatrix}~3\sqrt{2} q^{2/3} (1 + q^2 + 2 q^4 + 2 q^8 + q^{10}  +\cdots) ~\\~-(1 + 6 q^2 + 6 q^6 + 6 q^8  + \cdots) ~\end{pmatrix}\;,\\
	Y_{\rep{2'}II}^{(1)}&=\begin{pmatrix}-\sqrt{6}q^{1/6}(1 + 2 q + 2 q^2 + 2 q^3 + q^4 + 2 q^5 + 2 q^6 + 2 q^7 + 3 q^8 + 2 q^{10} + \cdots) \\
										~2\sqrt{3}q^{1/2} (1 + q + 2 q^3 + q^4 + 2 q^6 + 2 q^9 + 2 q^{10}+  \cdots) ~\end{pmatrix}\;.
	\end{align}
\end{subequations}

Differences are evident when comparing this case and that of the groups $\Gamma'_{3}$ and $2T$. 
For $\Gamma(2)/\Gamma(6)$, we see that \acp{VVMF} building doublets $\rep{2}'$ and $\rep{2}''$ of the finite modular group have minimal weight 1. 
Meanwhile, in the cases discussed in \Cref{sec:TprimeFromGamma} we find higher minimal modular weights. 
Furthermore, there are two different \acp{VVMF} of minimal weight building $\rep{2}'$ \acp{irrep} of $\Gamma(2)/\Gamma(6)$; instead, only one \ac{VVMF} transforming as $\rep2'$ appears in the $T'$ realizations discussed before.

\begin{table}[t!]
	{\centering
		\begin{tabular}{ccc}
			\toprule
			$\rep{r}$   & $\rho_{\Gamma(2) / \Gamma(6)}(a)$ & $\rho_{\Gamma(2) / \Gamma(6)}(b)$ \\ 
			\midrule
			$\rep{1}$   & $1$ & $1$ \\ 
			$\rep{1}'$  & $1$ & $\omega$ \\ 
			$\rep{1}''$ & $1$ & $\omega^2$ \\ 
			$\rep{2}$   & $\frac{\ii}{\sqrt3}
			\begin{pmatrix}
				1 & \sqrt 2  \\
				\sqrt 2 & -1
			\end{pmatrix}$ &
			$
			\omega \begin{pmatrix}
				1 & 0 \\
				0 & \omega
			\end{pmatrix}$ \\ 
			$\rep{2}'$ & $\frac{\ii}{\sqrt 3}
			\begin{pmatrix}
				1 & \sqrt 2 \\
				\sqrt 2  & -1
			\end{pmatrix}$ &
			$
			\omega^2 \begin{pmatrix}
				1 & 0 \\
				0 & \omega
			\end{pmatrix}$ \\ 
			$\rep{2}''$ & $\frac{\ii}{\sqrt 3}
			\begin{pmatrix}
				1 & \sqrt 2 \\
				\sqrt 2 & -1
			\end{pmatrix}$ &
			$
			\begin{pmatrix}
				1 & 0 \\
				0 & \omega
			\end{pmatrix}$ \\ 
			$\rep{3}$ & $\frac{1}{3}
			\begin{pmatrix}
				-1 & 2 & 2 \\
				2 & -1 & 2 \\
				2 & 2 & -1
			\end{pmatrix}$ &
			$
			\begin{pmatrix}
				1 & 0 & 0\\
				0 & \omega & 0\\
				0 & 0 & \omega^2
			\end{pmatrix}$ \\
			\bottomrule
		\end{tabular}
		\caption{\label{tab:Quotient_reps}
		Representations of the finite modular group $\Gamma(2) / \Gamma(6)\cong[24,3]$ in terms of the generators $a$ and $b$ of \Cref{eq:Gamma2overGamma6Presentation}, with $\omega=\ee^{2\pi \ii / 3}$. Notice that all \acp{irrep} are the same as in $\Gamma'_3$, cf.~\Cref{tab:Tprime_reps}, with the difference that the generators have a different image in $\Gamma(2) \subset \SL{2,\mathds{Z}}$.}
	}
\end{table}

It is well-known that the space of weight-2 modular forms for $\Gamma(6)$ can be generated by products of weight-1 modular forms. The dimension of this space is 12~\cite{cohen2017modular}, while there are 21 distinct and nonvanishing ways to combine the weight-1 modular forms in \Cref{eq:weight1FormsG6} to obtain weight-2 \acp{VVMF}.
This observation suggests that there must be some non-trivial algebraic relations between the six weight-1 modular forms that generate the entire space. One such relation is $Y_{6} Y_{4} = \left( Y_{1} Y_{5} +Y_{2} Y_{4} \right)/2$. All other algebraic relations can be found in \Cref{eq:G6relations} of~\Cref{app:table_expansions}.
Taking these algebraic relations into consideration, we can list all the modular form multiplets of weight 2 that are linearly independent by applying the tensor product rules in \Cref{eq:Tprime_tensorProducts}, leading to
\begin{subequations}\label{eq:Contractions}
	\begin{align}
		Y_{\rep{1}I}^{(2)} &= \left( Y^{(1)}_{\rep{2'}I} Y^{(1)}_{\rep{2''}} \right)_{\rep{1}}\;,&
		Y_{\rep{1}II}^{(2)} &= \left( Y^{(1)}_{\rep{2'}II} Y^{(1)}_{\rep{2''}} \right)_{\rep{1}}\;,&
		Y_{\rep{1''}}^{(2)}&= \left( Y^{(1)}_{\rep{2'}I} Y^{(1)}_{\rep{2'}II} \right)_{\rep{1''}}\;,\\
		Y_{\rep{3}I}^{(2)} &= \left( Y^{(1)}_{\rep{2'}I} Y^{(1)}_{\rep{2''}} \right)_{\rep{3}}\;,&
		Y_{\rep{3}II}^{(2)} &= \left( Y^{(1)}_{\rep{2'}II} Y^{(1)}_{\rep{2''}} \right)_{\rep{3}}\;,&
		Y_{\rep{3}III}^{(2)} &= \left( Y^{(1)}_{\rep{2''}} Y^{(1)}_{\rep{2''}} \right)_{\rep{3}}\;.
	\end{align}
\end{subequations}
The linearly independent modular form multiplets of weight 3 are obtained analogously as
\begin{subequations}\label{eq:weight3G2G6}
	\begin{align}
	Y^{(3)}_{\rep{2}I} &= \left( Y^{(1)}_{\rep{2''}} Y^{(2)}_{\rep{3}I} \right)_{\rep{2}}\;,&
	Y^{(3)}_{\rep{2}II} &= \left( Y^{(1)}_{\rep{2''}} Y^{(2)}_{\rep{3}II} \right)_{\rep{2}}\;,&
	Y^{(3)}_{\rep{2'}I} &= \left( Y^{(1)}_{\rep{2''}} Y^{(2)}_{\rep{3}I}  \right)_{\rep{2'}}\;,\\
	Y^{(3)}_{\rep{2'}II} &= \left( Y^{(1)}_{\rep{2''}} Y^{(2)}_{\rep{3}II}  \right)_{\rep{2'}}\;,&
	Y^{(3)}_{\rep{2'}III} &= \left( Y^{(1)}_{\rep{2''}} Y^{(2)}_{\rep{3}III}  \right)_{\rep{2'}}\;,&
	Y^{(3)}_{\rep{2'}IV} &= \left( Y^{(1)}_{\rep{2''}} Y^{(2)}_{\rep{1''}} \right)_{\rep{2'}}\;,\\
	Y^{(3)}_{\rep{2''}I} &= \left( Y^{(1)}_{\rep{2''}} Y^{(2)}_{\rep{1}I}  \right)_{\rep{2''}}\;,&
	Y^{(3)}_{\rep{2''}II} &= \left( Y^{(1)}_{\rep{2''}} Y^{(2)}_{\rep{1}II} \right)_{\rep{2''}}\;,&
	Y^{(3)}_{\rep{2''}III} &= \left( Y^{(1)}_{\rep{2''}} Y^{(2)}_{\rep{3}I}  \right)_{\rep{2''}}\;.
	\end{align}
\end{subequations}

All the remaining possible contractions can be shown to be either zero or linear combinations of those shown here.
It is remarkable that this method reproduces the results obtained by using the \ac{MLDE} for the group $\SL{2,\mathds{Z}} / \Gamma(3)$.
For example, the exact same $q$-expansion for the triplet in \Cref{tab:GammaPrime3_modularForm} were found in~\cite{Liu:2019khw}.
All the $q$-expansions for these modular forms can be found in \Cref{app:table_expansions}.

\section{Model building with different \texorpdfstring{$T'$}{T(prime)} modular flavor symmetries}
\label{sec:models}

Let us now work out some quark models based on the different modular instances of $T'$ that we have introduced so far.
In order to make the comparison explicit, we use consistent universal assignments for modular weights and representations.
That is, when we compare the phenomenology emerging from the various modular $T'$ symmetries, we choose the universal representations $\rho^{(I)}$ and weights $k_I$ for all of our models given by
\begin{align}
\rho^{(u^c)} &: \rep{2}'' \oplus \rep{1}\;,  & k_{u^c_{1,2,3}} &= 3,3,2\;,  &\rho^{(d^c)}     &: \rep{3}\;,   & k_{d^c}     &= 2\;,\notag \\
\rho^{(Q)}   &: \rep{3}\;,                      & k_{Q}           &= 1\;,         &\rho^{(H_{u/d})} &: \rep{1}\;,      & k_{H_{u/d}} &= -1\;.\label{eq:Model32RepWeight}
\end{align}
We denote the up-type quarks transforming as $\rep{2}''$ of $T'$ by $u^c_D\defeq(u^c_1,u^c_2)^\mathrm{T}$, and $u^c_3$ builds a trivial singlet under $T'$.
Note that, since the only differences in the representations given in \Cref{tab:Tprime_reps,tab:Quotient_reps} are found in the doublet representations, we naturally expect phenomenological differences in the up-type quark sector when exploring the various instances of modular $T'$.

\subsection{Model IA based on \texorpdfstring{$\Gamma'_3=\SL{2,\mathds{Z}}/\Gamma(3)$}{Gamma(3,prime)=SL(2,Z)/Gamma(3)}}
\label{sec:ModelIA}

For the group $\Gamma'_3 \cong T'$, with the charge assignments in \Cref{eq:Model32RepWeight}, the only possible \acp{VVMF} that appear in the superpotential at lowest order are $Y_{\rep2'}^{(3)}$, $Y_{\rep2''}^{(3)}$ of weight 3, and $Y_{\rep{3}}^{(2)}$ of weight 2.
The renormalizable,\footnote{Here, renormalizable means that the effective couplings after treating $\tau$ as a background field have mass dimensions no larger than 3.} modular invariant superpotential reads
	\begin{align}
		\mathscr{W}&\supset
		\left[
			\alpha_1 Y_{\rep2'}^{(3)} (u^c_D Q)_{\rep2''} + \alpha_2 Y_{\rep2''}^{(3)} (u^c_D Q)_{\rep2'}  + \alpha_3 Y_{\rep{3}}^{(2)} (u_3^c Q )_{\rep{3}}
		\right]_{\rep{1}} H_u \notag\\
		&\quad{}+ \left[
			\beta_1Y_{\rep{3}}^{(2)}  \left( d^c Q\right)_{\rep{3}_\mathrm{S}} +\beta_2 Y_{\rep{3}}^{(2)} \left(d^c Q \right)_{\rep{3}_\mathrm{A}}
		\right]_{\rep{1}} H_d\;.\label{eq:TpW}
	\end{align}
Here, the subscripts in $\rep{3}_\mathrm{S}$ and $\rep{3}_\mathrm{A}$ indicate respectively the symmetric and antisymmetric contractions, as defined in \Cref{eq:Tprime_tensorProducts}. 
As is common practice in bottom-up model building, we introduced coefficients $\alpha_i$ and $\beta_i$, which may be used to fit data.
The superpotential~\eqref{eq:TpW} gives rise to the up and down-quark mass matrices
\begin{subequations}
	\label{eq:MassMatricesTp}
	\begin{align}
		M_u &=
		\begin{pmatrix}
			-\alpha_1 Y_{\rep2',2}^{(3)}-\alpha_2 \sqrt{2}Y_{\rep2'',1}^{(3)} & \alpha_1 \sqrt{2}Y_{\rep2',1}^{(3)} & \alpha_2
			Y_{\rep2'',2}^{(3)} \\
		  -\alpha_1 Y_{\rep2',1}^{(3)} & \alpha_2 \sqrt{2} Y_{\rep2'',2}^{(3)} & \alpha_2 Y_{\rep2'',1}^{(3)}-\alpha_1 \sqrt{2}
			Y_{\rep2',2}^{(3)} \\
		  \alpha_3 Y_{\rep{3},1}^{(2)} & \alpha_3 Y_{\rep{3},3}^{(2)} & \alpha_3 Y_{\rep{3},2}^{(2)} \\
		\end{pmatrix}\,
		v_u\;,\\
\label{eq:MdModelI}
		M_d &=
		\begin{pmatrix}
			2 \beta_1 Y_{\rep{3},1}^{(2)} & \beta_2 Y_{\rep{3},3}^{(2)}-\beta_1 Y_{\rep{3},3}^{(2)} & -\beta_1 Y_{\rep{3},2}^{(2)}-\beta_2 Y_{\rep{3},2}^{(2)} \\
			-\beta_1 Y_{\rep{3},3}^{(2)}-\beta_2 Y_{\rep{3},3}^{(2)} & 2 \beta_1 Y_{\rep{3},2}^{(2)} & \beta_2 Y_{\rep{3},1}^{(2)}-\beta_1 Y_{\rep{3},1}^{(2)} \\
			\beta_2 Y_{\rep{3},2}^{(2)}-\beta_1 Y_{\rep{3},2}^{(2)} & -\beta_1 Y_{\rep{3},1}^{(2)}-\beta_2 Y_{\rep{3},1}^{(2)} & 2 \beta_1 Y_{\rep{3},3}^{(2)} \\
		   \end{pmatrix}\,
		v_d \;,
	\end{align}
\end{subequations}
where $v_{u/d}$ stand for the \ac{VEV} of the Higgs fields $H_{u/d}$, and the subscripts of the modular forms $Y$, after the comma, indicate the components of the \ac{VVMF}.

The  determinants of the mass matrices are given by
\begin{subequations}
\begin{align}
  \det(M_{u})&=\frac{4}{81}\sqrt{2} \bigl(\alpha_1- 108 \alpha_2\bigr)\, \bigl(\alpha_1 + 54 \alpha_2\bigr)\, \alpha_3\,\eta^{16}(\tau)\;,\label{eq:determinantMu_IA}\\
  \det(M_{d})&=-2 \beta_{1}\left(\beta^{2}_{1}-\beta^{2}_{2} \right) E_{6}(\tau)\;.\label{eq:determinantMd_allModels}
\end{align}
\end{subequations}
We observe that they are 1-dimensional \acp{VVMF} on the modular group $\SL{2,\mathds{Z}}$, which is to be expected because, in general, the determinant of the mass matrices of a model based on a finite modular group $\Upsilon/G_d$ must be a 1-dimensional modular form of the original modular group $\Upsilon$~\cite{Liu:2021gwa}.
In addition, one can study the phenomenology of our models close to critical points, such as $\tau=\ii,\omega$ or $\ii\infty$, where successful properties can emerge~\cite{Feruglio:2021dte,Feruglio:2022koo,Feruglio:2023mii,Petcov:2022fjf}. The near-critical behavior of the up-quark masses in the vicinity of $\tau\to\ii\infty$ is
\begin{equation}
\label{eq:massHierarchyIA}
m_u:~m_c:~m_t ~\approx~ q_3^2~:~ 1~ :~ 1\;,
\end{equation}
where $q_3\defeq q^{1/3}=\ee^{2\pi\ii\tau/3}$. 
Recall that the behavior indicated this way is implied as a relation in order of magnitude.
On the other hand, since $E_6(\tau)$ has a zero at $\tau=\ii$, for the down-quark sector the mass hierarchies near the vicinity of $\tau=\ii$ follow the scheme
\begin{equation}
\label{eq:massHierarchyIA2}
m_d:~m_s:~m_b ~\approx~ \varepsilon~:~ 1 ~:~ 1\;,
\end{equation}
where $\varepsilon=\frac{\tau-\ii}{\tau+\ii}$ is small. 
These mass relations indicate that, in this model, mass hierarchies cannot be explained through properties of the modular forms.
While there is the freedom to obtain hierarchies by carefully adjusting the coefficients $\alpha_i$ and $\beta_i$ in \Cref{eq:determinantMu_IA,eq:determinantMd_allModels}, we can conclude that this model is less promising than what one could have hoped for since the hierarchies have to be put in by hand.
As we shall see next in \Cref{sec:ModelIIA}, switching to ``another modular $T'$'' symmetry improves the situation considerably.

\subsection{Model IIA based on \texorpdfstring{$2T=\SL{2,\mathds{Z}}/N_{[24,3]II}$}{2T=SL(2,Z)/N([24,3]II)}}
\label{sec:ModelIIA}

For the group $2T \cong T'$, we consider again the representations and modular weights of the matter fields shown in~\Cref{eq:Model32RepWeight}. The differences can only arise from the fact that we must now consider the $2T$ modular forms given in \Cref{tab:2T_modularForm}. The only $2T$ \acp{VVMF} that appear in the superpotential at lowest order are $Y_{\rep2'}^{(3)}$ and $Y_{\rep{3}}^{(2)}$. For the up quark sector, the renormalizable, modular invariant superpotential is given by
\begin{equation}
	\label{eq:2TW}
		\mathscr{W} \supset
		\left[
			\alpha_1 Y_{\rep2'}^{(3)} (u^c_D Q)_{\rep2''}  + \alpha_2 Y_{\rep{3}}^{(2)} (u_3^c Q)_{\rep{3}}
		\right]_{\rep{1}} H_u  + \left[
			\beta_1 Y_{\rep{3}}^{(2)} \left( d^c Q\right)_{\rep{3}_\mathrm{S}} +\beta_2 Y_{\rep{3}}^{(2)} \left(d^c Q \right)_{\rep{3}_\mathrm{A}}
		\right]_{\rep{1}} H_d\;.
\end{equation}
Note that the superpotential for the down-quark sector coincides with the corresponding part of the previous model, cf.~\Cref{eq:TpW}. If then follows that the mass matrices can be expressed as
\begin{subequations}
\label{eq:MassMatrices2T}
\begin{align}
	M_u &=
		\begin{pmatrix}
			-\alpha_1 Y_{\rep2',2}^{(3)} & \alpha_1 \sqrt{2} Y_{\rep2',1}^{(3)} & 0 \\
			-\alpha_1 Y_{\rep2',1}^{(3)} & 0 & -\alpha_1 \sqrt{2} Y_{\rep2',2}^{(3)} \\
			\alpha_2 Y_{\rep{3},1}^{(2)} & \alpha_2 Y_{\rep{3},3}^{(2)} & \alpha_2 Y_{\rep{3},2}^{(2)} \\
		\end{pmatrix}
		\,v_u\;, \\
\label{eq:MdModelII}
	M_d &=
		\begin{pmatrix}
			2 \beta_1 Y_{\rep{3},1}^{(2)} & \beta_2 Y_{\rep{3},3}^{(2)}-\beta_1 Y_{\rep{3},3}^{(2)} & -\beta_1 Y_{\rep{3},2}^{(2)}-\beta_2 Y_{\rep{3},2}^{(2)} \\
			-\beta_1 Y_{\rep{3},3}^{(2)}-\beta_2 Y_{\rep{3},3}^{(2)} & 2 \beta_1 Y_{\rep{3},2}^{(2)} & \beta_2 Y_{\rep{3},1}^{(2)}-\beta_1 Y_{\rep{3},1}^{(2)} \\
			\beta_2 Y_{\rep{3},2}^{(2)}-\beta_1 Y_{\rep{3},2}^{(2)} & -\beta_1 Y_{\rep{3},1}^{(2)}-\beta_2 Y_{\rep{3},1}^{(2)} & 2 \beta_1 Y_{\rep{3},3}^{(2)} \\
		   \end{pmatrix}
		\,v_d \;,
\end{align}
\end{subequations}
where, for completeness, we give explicitly $M_d$ here even though it coincides with the matrix~\eqref{eq:MdModelI} of the previous model.
The determinant of the mass matrix $M_{u}$ is
\begin{equation}
\label{eq:determinantMu_IIA}
   \det(M_{u})= -12\sqrt{2}  \alpha^{2}_{1}\alpha_{2} \eta^{16}(\tau)\;,
\end{equation}
and the determinant of the matrix $M_{d}$ is given again by \Cref{eq:determinantMd_allModels}.
In addition, the near-critical behavior of the up-quark masses in the vicinity of $\tau\to\ii\infty$ reads
\begin{equation}
\label{eq:massHierarchyIIA}
   m_u:~m_c:~m_t ~\approx~ q_6^{3}:~ q_6 :~ 1\;,
\end{equation}
where $q_6\defeq q^{1/6}=\ee^{\pi\ii\tau/3}$. 
If one identifies $q_6$ with $\varepsilon^2$ or $\varepsilon^3$, where $\varepsilon$ denotes the hierarchy of a Froggatt--Nielsen~\cite{Froggatt:1978nt} (FN) model, then the hierarchy \eqref{eq:massHierarchyIIA} is rather compatible with what is being used in FN model building (cf.\ e.g.\ \cite{Fedele:2020fvh}).

In the group $2T$, there is only one possible modular form of weight 3 we can use, namely $Y_{\rep2'}^{(3)}$. This results in a zero texture in the mass matrix $M_u$, leading to distinct phenomenology compared to the one found for the group $\Gamma'_3$. On the other hand, the triplet modular forms are the same in both $T'$ and $2T$. As a result, the terms contracted with $Y_{\rep{3}}^{(3)}$ are identical in both the $2T$ and $T'$ models, leading to the same mass matrix for the down-quark sector.
In addition, it should be emphasized that since the $\rho_{2T}(T)$ has order 6, instead of order 3 as in $\Gamma_3'$, the near-critical behavior of the corresponding model near $\tau\to\ii\infty$ is more likely to produce large mass hierarchies, as has been shown by~\Cref{eq:massHierarchyIIA}. 
Therefore, the model based on $2T$ has far better prospects of describing the real world than its $\Gamma'_3$-based cousin.
The detailed analysis of quark models based on $2T$ are left for the future.

\subsection{Model IIIA based on \texorpdfstring{$\Gamma(2)/\Gamma(6)$}{Gamma(2)/Gamma(6)}}
\label{sec:ModelIIIA}

We now study a model based on the modular flavor symmetry $\Gamma(2)/\Gamma(6)$ with the universal assignments of representations and weights for the matter fields given in~\Cref{eq:Model32RepWeight}. In this case, we have a larger number of modular forms available, as listed in~\Cref{eq:Contractions,eq:weight3G2G6}, consisting in
2 modular forms in weight 3 in $\rep{2}$,
4 modular forms in weight 3 in $\rep{2}'$,
3 modular forms in weight 3 in $\rep{2}''$,
3 modular forms in weight 2 in $\rep{3}$,
2 modular forms in weight 2 in $\rep{1}$, and
1 modular forms in weight 2 in $\rep{1}''$, cf.~\Cref{sec:TprimeFromGamma2}.

The renormalizable superpotential is somewhat similar to the one found in the previous $T'$ models, but has more terms due to the availability of additional modular forms.
Specifically, for the up-quark sector, we use nine different doublet \acp{VVMF} and three different triplet \acp{VVMF}. 
For the down-quark sector, we use three different triplet \acp{VVMF} and three singlet modular forms.
We find
\begin{align}
		\mathscr{W}&\supset
		\left[
			\mathcal{Y}^{(3)}_{\rep2'} (u^c_D Q)_{\rep2''} + \mathcal{Y}^{(3)}_{\rep2''} (u^c_D Q)_{\rep2'} + \mathcal{Y}^{(2)}_{\rep{3}} (u_3^c Q)_{\rep{3}}
		\right]_{\rep{1}} H_u \notag\\
		&\quad{}+ \left[
			\mathcal{Y}^{(2)}_{\rep{3}}  \left(d^c Q\right)_{\rep{3}_\text{S}} + \mathcal{X}^{(2)}_{\rep{3}} \left(d^c Q \right)_{\rep{3}_\text{A}}+ \mathcal{Y}^{(2)}_{\rep{1'}} (d^c Q)_{\rep{1''}} + \mathcal{Y}^{(2)}_{\rep{1}} (d^c Q)_{\rep{1}}
		\right]_{\rep{1}} H_d\;.\label{eq:G2RG6}
\end{align}
The modular coefficients $\mathcal{Y}_{\rep{r}}^{(k)}$ and $\mathcal{X}_{\rep{r}}^{(k)}$ denote generic modular-form multiplets in the representation $\rep{r}$ with weight $k$, which are linear combinations of the linearly independent \acp{VVMF} in the same representation and weight: $\mathcal{Y}_{\rep{r}}^{(k)} = \sum_I a_I Y_{\rep{r}I}^{(k)}$.
For the up-quark sector, we have
\begin{subequations}
\label{eq:G2RG6Up}
\begin{align}
\mathcal Y_{\rep2''}^{(3)}&=\alpha_8 Y_{\rep2''I}^{(3)}+\alpha_9 Y_{\rep2''{II}}^{(3)}+\alpha_{10} Y_{\rep2''{III}}^{(3)}\;,\\
\mathcal Y_{\rep2'}^{(3)} &= \alpha_4 Y_{\rep2'I}^{(3)}+\alpha_5 Y_{\rep2'{II}}^{(3)}+\alpha_6 Y_{\rep2'{III}}^{(3)}+\alpha_7 Y_{\rep2'{IV}}^{(3)}\;,\\
\mathcal Y_{\rep{2}}^{(3)} &= \alpha_{11} Y_{\rep{2}I}^{(3)}+\alpha_{12} Y_{\rep{2}{II}}^{(3)}\;,\\
\mathcal Y_{\rep{3}}^{(2)} &= \alpha_1 Y_{\rep{3}I}^{(2)}+\alpha_2 Y_{\rep{3}{II}}^{(2)}+\alpha_3 Y_{\rep{3}{III}}^{(2)}\;.
\end{align}
\end{subequations}
For the down-quark sector, we find
\begin{subequations}
\label{eq:G2RG6Down}
\begin{align}
	\mathcal Y_{\rep{1}}^{(2)} &=  \beta_8 Y_{\rep{1}I}^{(2)}+\beta_9 Y_{\rep{1}{II}}^{(2)}\;,\\
	\mathcal Y_{\rep{1}''}^{(2)} &= \beta_7 Y_{\rep{1}''}^{(2)}\;,\\
	\mathcal Y_{\rep{3}}^{(2)} &= \beta_1 Y_{\rep{3}I}^{(2)}+\beta_3 Y_{\rep{3}{II}}^{(2)}+\beta_5 Y_{\rep{3}{III}}^{(2)}\;,\\
	\mathcal X_{\rep{3}}^{(2)} &= \beta_2 Y_{\rep{3}I}^{(2)}+\beta_4 Y_{\rep{3}{II}}^{(2)}+\beta_6 Y_{\rep{3}{III}}^{(2)}\;.
\end{align}
\end{subequations}
There are 12 different coefficients $\alpha_i$ in the up-quark sector and 9 different coefficients $\beta_i$ in the down-quark sector.
This model, therefore, has a total of 21 free parameters of this type.
Consequently, the quark mass matrices of this model are found to be
\begin{subequations}\label{eq:MassMatricesG2RG6}
	\begin{align}
		M_u=&
		\begin{pmatrix}
			-\sqrt{2} \mathcal Y_{\rep2'',1}^{(3)} - \mathcal Y_{\rep2',2}^{(3)} & \sqrt{2} \mathcal Y_{\rep2',1}^{(3)}+ \mathcal Y_{\rep{2},2}^{(3)} & -\sqrt{2} \mathcal Y_{\rep{2},1}^{(3)} + \mathcal Y_{\rep2'',2}^{(3)}\\
			-\mathcal Y_{\rep2',1}^{(3)} +\sqrt{2} \mathcal Y_{\rep{2},2}^{(3)} & \mathcal Y_{\rep2'',2}^{(3)}+\mathcal Y_{\rep{2},1}^{(3)} & -\sqrt{2} \mathcal Y_{\rep2',2}^{(3)}+\mathcal Y_{\rep2'',1}^{(3)} \\
			\sqrt{2} \mathcal Y_{\rep{3},1}^{(2)} & \sqrt{2} \mathcal Y_{\rep{3},3}^{(2)} & \sqrt{2} \mathcal Y_{\rep{3},2}^{(2)} \\
		\end{pmatrix}
		v_u\;,\\
		M_d=&
		\begin{pmatrix}
			\sqrt{2} \mathcal Y_{\rep{1},1}^{(2)} +2 \mathcal Y_{\rep{3},1}^{(2)} & \sqrt{2} \mathcal Y_{\rep{1}'',1}^{(2)} - \mathcal Y_{\rep{3},3}^{(2)}+ \mathcal X_{\rep{3},3}^{(2)} & -\mathcal Y_{\rep{3},2}^{(2)}-\mathcal X_{\rep{3},2}^{(2)} \\
			\sqrt{2} \mathcal Y_{\rep{1}'',1}^{(2)} - \mathcal Y_{\rep{3},3}^{(2)} - \mathcal X_{\rep{3},3}^{(2)}  & 2 \mathcal Y_{\rep{3},2}^{(2)} & \sqrt{2} \mathcal Y_{\rep{1},1}^{(2)} -  \mathcal Y_{\rep{3},1}^{(2)}+ \mathcal X_{\rep{3},1}^{(2)} \\
			-\mathcal Y_{\rep{3},2}^{(2)} + \mathcal X_{\rep{3},2}^{(2)} & \sqrt{2} \mathcal Y_{\rep{1},1}^{(2)} -\mathcal X_{\rep{3},1}^{(2)} - \mathcal Y_{\rep{3},1}^{(2)} & \sqrt{2} \mathcal Y_{\rep{1}'',1}^{(2)}+ 2 \mathcal Y_{\rep{3},3}^{(2)} \\
		   \end{pmatrix}
		v_d \;,
	\end{align}
\end{subequations}
where $v_{u/d}$ stands for the \ac{VEV} of the Higgs fields $H_{u/d}$, and the subscripts of the modular forms $Y$ after the comma indicate the components of the \ac{VVMF}.

As before, one can compute the determinant of the mass matrices, which turn out to be quite complicated and not so enlightening for our discussion. However, we still see that they are 1-dimensional modular forms of $\Gamma(2)$, which cannot be expressed simply in terms of $\eta(\tau)$, $E_4(\tau)$ and $E_6(\tau)$. For the up-quark sector, the leading term of the $q$-expansion is proportional to $q^{1/6}$, indicating that some hierarchies might be achievable close to $\tau\to\ii\infty$. The near-critical behavior of the up-quark masses in the vicinity of this critical point is given by
\begin{equation}
\label{eq:massHierarchyIIIA}
	m_u ~:~ m_c ~:~ m_t ~\approx~ q_6 ~:~ 1 ~:~ 1 \;,
\end{equation}
where $q_6 \defeq q^{1/6} = \ee^{\pi\ii\tau/3}$. As $\tau\to\ii\infty$, the down-quark masses behave as
\begin{equation}
    m_d ~:~ m_s ~:~ m_b ~\approx~ 1 ~:~ 1 ~:~ 1\;,
\end{equation}
displaying a very different scaling in comparison with the models based on $\Gamma'_3$ and $2T$, cf.~\Cref{eq:massHierarchyIA2}. As we shall shortly show in \Cref{sec:3TprimeSummary}, this does not prevent us from finding a large imaginary value of $\tau$ to fit the model to observations due to the large number of parameters appearing in models based on $\Gamma(2)/\Gamma(6)$.

\subsection{Phenomenological features of three instances of \texorpdfstring{$T'$}{T(prime)}}
\label{sec:3TprimeSummary}

Let us now analyze the main phenomenological features of flavor models based on the different instances of modular $T'$ flavor symmetry studied in this work, $\Gamma_3'$, $2T$ and $\Gamma(2)/\Gamma(6)$. 
We have called these models IA, IIA and IIIA, where the letter A corresponds to the weight and representation assignments of \Cref{eq:Model32RepWeight}. 
We have summarized the results of \Cref{sec:ModelIA,sec:ModelIIA,sec:ModelIIIA} in \Cref{tab:ModelsummaryA}.

\begin{table}[t!]
\centering
\begin{tabular}{ccccccccccccc}\toprule
IA & $u^c$ & $d^c$ & $Q$ & $H_{u/d}$ & $Y^{(3)}_{\rep2''}$ & $Y^{(3)}_{\rep2'}$ & $Y^{(2)}_{\rep{3}}$&
\multicolumn{1}{c}{}  &\multicolumn{1}{c}{}  &\multicolumn{1}{c}{}  &\\ 
\midrule
$\rho_{\Gamma_3'}$ & $\rep2'' \oplus \rep{1}$ & $\rep{3}$ & $\rep{3}$ & $\rep{1}$ & $\rep2''$ & $\rep2'$ & $\rep{3}$&
\multicolumn{1}{c}{}  &\multicolumn{1}{c}{}  &\multicolumn{1}{c}{}  & \\ 
$k$ & $(3,2)$ & $2$ & $1$ & $-1$ & $3$ & $3$ & $2$ &
\multicolumn{1}{c}{}  &\multicolumn{1}{c}{}  &\multicolumn{1}{c}{}  & \\ 
\# of $Y$ & $-$ & $-$ & $-$ & $-$ & 1 & 1 & 1 &
\multicolumn{1}{c}{}  &\multicolumn{1}{c}{}  &\multicolumn{1}{c}{}  & \\ 
$M_u$ & \multicolumn{11}{c}{{\small
$\begin{aligned}
\begin{pmatrix}
-\alpha_1 Y_{\rep2',2}^{(3)}-\alpha_2 \sqrt{2} Y_{\rep2'',1}^{(3)} & \alpha_1 \sqrt{2} Y_{\rep2',1}^{(3)} &\alpha_2
Y_{\rep2'',2}^{(3)} \\
-\alpha_1 Y_{\rep2',1}^{(3)}& \alpha_2  \sqrt{2} Y_{\rep2'',2}^{(3)} & \alpha_2 Y_{\rep2'',1}^{(3)}-\alpha_1 \sqrt{2}
Y_{\rep2',2}^{(3)} \\
 \alpha_3 Y_{\rep{3},1}^{(2)} & \alpha_3 Y_{\rep{3},3}^{(2)} & \alpha_3 Y_{\rep{3},2}^{(2)} \\
\end{pmatrix}v_u
\end{aligned}$} } \\ 
$M_d$ & \multicolumn{11}{c}{{\small
$\begin{aligned}	\begin{pmatrix}
2 \beta_1 Y_{\rep{3},1}^{(2)} & \beta_2 Y_{\rep{3},3}^{(2)}-\beta_1 Y_{\rep{3},3}^{(2)} & -\beta_1 Y_{\rep{3},2}^{(2)}-\beta_2 Y_{\rep{3},2}^{(2)} \\
-\beta_1 Y_{\rep{3},3}^{(2)}-\beta_2 Y_{\rep{3},3}^{(2)} & 2 \beta_1 Y_{\rep{3},2}^{(2)} & \beta_2 Y_{\rep{3},1}^{(2)}-\beta_1 Y_{\rep{3},1}^{(2)} \\
\beta_2 Y_{\rep{3},2}^{(2)}-\beta_1 Y_{\rep{3},2}^{(2)} & -\beta_1 Y_{\rep{3},1}^{(2)}-\beta_2 Y_{\rep{3},1}^{(2)} & 2 \beta_1 Y_{\rep{3},3}^{(2)} \\
\end{pmatrix}v_d \end{aligned}$} } \\   
\midrule
IIA & $u^c$ & $d^c$ & $Q$ & $H_{u/d}$ & $Y^{(3)}_{\rep2'}$ & $Y^{(2)}_{\rep{3}}$ &
\multicolumn{1}{c}{}  &\multicolumn{1}{c}{}  &\multicolumn{1}{c}{}  &\multicolumn{1}{c}{}  &  \\ 
$\rho_{2T}$ & $\rep2'' \oplus \rep{1}$ & $\rep{3}$ & $\rep{3}$ & $\rep{1}$ & $\rep2'$ & $\rep{3}$&
\multicolumn{1}{c}{}  &\multicolumn{1}{c}{}  &\multicolumn{1}{c}{}  &\multicolumn{1}{c}{}  & \\ 
$k$ & $(3,2)$ & $2$ & $1$ & $-1$ & $3$ & $2$ &
\multicolumn{1}{c}{}  &\multicolumn{1}{c}{}  &\multicolumn{1}{c}{}  &\multicolumn{1}{c}{}  & \\ 
\# of $Y$ & $-$ & $-$ & $-$ & $-$ & 1 & 1 &
\multicolumn{1}{c}{}  &\multicolumn{1}{c}{}  &\multicolumn{1}{c}{}  &\multicolumn{1}{c}{}  & \\ 
$M_u$ & \multicolumn{11}{c}{{\small
$\begin{aligned}
\begin{pmatrix}
-\alpha_1 Y_{\rep2',2}^{(3)} & \alpha_1 \sqrt{2} Y_{\rep2',1}^{(3)} & 0 \\
-\alpha_1 Y_{\rep2',1}^{(3)} & 0 & -\alpha_1 \sqrt{2} Y_{\rep2',2}^{(3)} \\
\alpha_2 Y_{\rep{3},1}^{(2)} & \alpha_2 Y_{\rep{3},3}^{(2)} & \alpha_2 Y_{\rep{3},2}^{(2)} \\
\end{pmatrix}v_u
\end{aligned}$} } \\ 
$M_d$ & \multicolumn{11}{c}{{\small
$\begin{aligned}
\begin{pmatrix}
2 \beta_1 Y_{\rep{3},1}^{(2)} & \beta_2 Y_{\rep{3},3}^{(2)}-\beta_1 Y_{\rep{3},3}^{(2)} & -\beta_1 Y_{\rep{3},2}^{(2)}-\beta_2 Y_{\rep{3},2}^{(2)} \\
-\beta_1 Y_{\rep{3},3}^{(2)}-\beta_2 Y_{\rep{3},3}^{(2)} & 2 \beta_1 Y_{\rep{3},2}^{(2)} & \beta_2 Y_{\rep{3},1}^{(2)}-\beta_1 Y_{\rep{3},1}^{(2)} \\
\beta_2 Y_{\rep{3},2}^{(2)}-\beta_1 Y_{\rep{3},2}^{(2)} & -\beta_1 Y_{\rep{3},1}^{(2)}-\beta_2 Y_{\rep{3},1}^{(2)} & 2 \beta_1 Y_{\rep{3},3}^{(2)} \\
\end{pmatrix}v_d \end{aligned}$} } \\   
\midrule
IIIA& $u^c$ & $d^c$ & $Q$ & $H_{u/d}$ & $Y^{(3)}_{\rep{2}}$ & $Y^{(3)}_{\rep2'}$ & $Y^{(3)}_{\rep2''}$ & $Y^{(2)}_{\rep{3}}$  & $Y^{(2)}_{\rep{1}}$& $Y^{(2)}_{\rep{1''}}$ & \\ 
$\rho_{\Gamma(2)/\Gamma(6)}$ & $\rep2'' \oplus \rep{1}$ & $\rep{3}$ & $\rep{3}$ & $\rep{1}$ & $\rep{2}$ & $\rep2'$ & $\rep2''$ & $\rep{3}$ & $\rep{1}$ & $\rep{1''}$ &\\ 
$k$ & $(3,2)$ & $2$ & $1$ & $-1$ & $3$ & $3$ & $3$ & $2$ & $2$ & $2$ &\\ 
\# of $Y$ & $-$ & $-$ & $-$ & $-$ & 2 & 4 & 3 & 3 & 2 & 1 & \\ 
$M_u$ & \multicolumn{11}{c}{{\small
$\begin{aligned}
\begin{pmatrix}
-\sqrt{2} \mathcal{Y}_{\rep2'',1}^{(3)}-\mathcal{Y}_{\rep2',2}^{(3)} & \sqrt{2} \mathcal{Y}_{\rep2',1}^{(3)} + \mathcal{Y}_{\rep{2},2}^{(3)} & \mathcal{Y}_{\rep2'',2}^{(3)}-\sqrt{2} \mathcal{Y}_{\rep{2},1}^{(3)} \\
\sqrt{2} \mathcal{Y}_{\rep{2},2}^{(3)}-\mathcal{Y}_{\rep2',1}^{(3)} & \sqrt{2} \mathcal{Y}_{\rep2'',2}^{(3)}+ \mathcal{Y}_{\rep{2},1}^{(3)} & \mathcal{Y}_{\rep2'',1}^{(3)}-\sqrt{2} \mathcal{Y}_{\rep2',2}^{(3)} \\
\sqrt{2} \mathcal{Y}_{\rep{3},1}^{(2)} & \sqrt{2} \mathcal{Y}_{\rep{3},3}^{(2)} & \sqrt{2} \mathcal{Y}_{\rep{3},2}^{(2)} \\
\end{pmatrix}v_u
\end{aligned}$} } \\ 
$M_d$ & \multicolumn{11}{c}{{\small
$\begin{aligned}\begin{pmatrix}
\sqrt{2}\mathcal{Y}_{\rep{1}}^{(2)}+2 \mathcal{Y}_{\rep{3},1}^{(2)} & \sqrt{2} \mathcal{Y}_{\rep{1}''}^{(2)} - \mathcal{Y}_{\rep{3},3}^{(2)} + \mathcal{X}_{\rep{3},3}^{(2)}  & -\mathcal{Y}_{\rep{3},2}^{(2)} -  \mathcal{X}_{\rep{3},2}^{(2)} \\
\sqrt{2} \mathcal{Y}_{\rep{1}''}^{(2)} - \mathcal{Y}_{\rep{3},3}^{(2)} -  \mathcal{X}_{\rep{3},3}^{(2)} & 2 \mathcal{Y}_{\rep{3},2}^{(2)} & \sqrt{2} \mathcal{Y}_{\rep{1}}^{(2)} -\mathcal{Y}_{\rep{3},1}^{(2)} + \mathcal{X}_{\rep{3},1}^{(2)} \\
-\mathcal{Y}_{\rep{3},2}^{(2)} + \mathcal{X}_{\rep{3},2}^{(2)}  & \sqrt{2} \mathcal{Y}_{\rep{1}}^{(2)} - \mathcal{Y}_{\rep{3},1}^{(2)} -  \mathcal{X}_{\rep{3},1}^{(2)} & \sqrt{2} \mathcal{Y}_{\rep{1}''}^{(2)} + 2 \mathcal{Y}_{\rep{3},3}^{(2)} \\
\end{pmatrix}v_d \end{aligned}$} } \\ \bottomrule
\end{tabular}
\caption{\label{tab:ModelsummaryA}
Summary of features of quark models IA based on $\Gamma'_3$, IIA based on $2T$, and IIIA based on $\Gamma(2)/\Gamma(6)$,
all with identical representation and charge assignments, as given in \Cref{eq:Model32RepWeight}.}
\end{table}

The main immediate difference is the different number of the various \acp{VVMF} appearing in the renormalizable superpotential of each instance of modular $T'$.\footnote{Not all basis \acp{VVMF} appear because of the choice of modular weights.} 
The impact of these differences is evident when comparing the up-quark mass matrices of models IA and IIA.
We see that the zeroes in $M_u$ appear in model IIA because there is no $Y^{(2)}_{\rep2''}(\tau)$ in this case. 
Apart from this difference, these two models are equivalent.
However, in model IIIA the number of available basis \acp{VVMF} is large in all $T'$ representations. 
In particular, the entries of $M_u$ that are zero in the model IIA contain additional contributions from the modular forms transforming as $\rep{2}$ (on top of those transforming as $\rep{2}''$, as in model IA). 
Further, $M_d$ also receives contributions from a comparatively large number of modular forms.
As a consequence, there are more coefficients $\alpha_i$ and $\beta_j$, which, in the context of bottom-up model building are free parameters.
In slightly more detail, while models IA and IIA have 6 of these parameters each, in model IIIA there are 21 of them.
An increase in the number of free parameters makes it generally easier to fit data but at the same time reduces the predictive power of a given model.

Another important difference, which arises as a consequence of the admissible modular forms in each case, is the near-critical behavior of the up-quark masses close to $\tau\to\ii\infty$. 
As shown in \Cref{eq:massHierarchyIA,eq:massHierarchyIIA,eq:massHierarchyIIIA}, very close to the cusp, model IIA is more suited in obtaining mass hierarchies from the properties of modular forms, as opposed to getting them from adjusting parameters.

Some remarks on the $T'$ realization based on $\Gamma(2)/\Gamma(6)$ are in order. 
For models endowed with this symmetry, it is important to note that there are no elliptic fixed points, i.e.\ fixed points inside $\mathcal{H}$, such as $\tau=\ii$ or $\omega$. 
Rather, only the cusps $0$, $1$, and $\ii\infty$ serve as the fixed points in the fundamental domain $\mathcal{F}_2$ of $\Gamma(2)$, see \Cref{fig:fundamentaldomain}. 
Therefore, we can anticipate that promising universal near-critical behavior, leading to hierarchies originating from properties of modular forms, will only exist for these cusps. 
This is clearly different from cases involving $\Gamma'_3$ and $2T$. 
Even when considering the near-critical behavior in the vicinity of $\tau\to\ii\infty$, it should be noted that the determinant $\det(M_u)$ of model IIIA remains a 1-dimensional \ac{VVMF} transforming as $\rep{1}''$ of $\Gamma(2)/\Gamma(6)$, but it is a \ac{VVMF} defined on $\Gamma(2)$ instead of $\mathrm{SL}(2,\mathds{Z})$. 
Consequently, it is no longer proportional to $\eta^{16}(\tau)$. 
As a result, the near-critical behavior of the determinants in model IIIA differs significantly from the ones observed in models IA and IIA.
A more detailed analysis of these questions is left for future works.

The final phenomenological test for our models should be the actual contrast between the model predictions and observations. 
Just as a quick verification to arrive at a side-remark on the phenomenological potential of our models, we have performed a scan to fit the parameters of all three models. 
Some relevant details are presented in \Cref{app:fittingResults}. As discussed there, models IA and IIA fail to reproduce observations. It is interesting that these models are based on quotients of $\SL{2,\mathds{Z}}$ and its normal subgroups, which models with modular flavor symmetries are traditionally based on.
On the other hand, a modular flavor model based on $\Gamma(2)/\Gamma(6)\cong T'$ can successfully fit data for large but finite $\im\tau$.
This adds more substance to our findings: choosing a finite modular flavor symmetry is not enough to achieve successful phenomenology. Rather, the origin of the finite symmetry matters.

\begin{table}[t!]
\centering
\begin{tabular}{cccccccccccc}\toprule
IB & $u^c$ & $d^c$ & $Q$ & $H_{u/d}$ & $Y^{(1)}_{\rep2''}$ & $Y^{(2)}_{\rep{3}}$&
\multicolumn{1}{c}{}  &\multicolumn{1}{c}{} &\multicolumn{1}{c}{} &  \\ 
$\rep{\rho}$ & $\rep2'' \oplus \rep{1}$ & $\rep{3}$ & $\rep{3}$ & $\rep{1}$ & $\rep2''$ & $\rep{3}$&
 \multicolumn{1}{c}{} & \multicolumn{1}{c}{} &\multicolumn{1}{c}{} & \\ 
$k$ & $(1,2)$ & $2$ & $1$ & $-1$ & $1$ & $2$ &
 \multicolumn{1}{c}{} &\multicolumn{1}{c}{} & \multicolumn{1}{c}{} &   \\ 
\# of $Y$ & $-$ & $-$ & $-$ & $-$ & 1  & 1 &
\multicolumn{1}{c}{}& \multicolumn{1}{c}{}& \multicolumn{1}{c}{} &   \\ 
$M_u$ & \multicolumn{10}{c}{{\small
$\begin{aligned}
\begin{pmatrix}
-\alpha_1 \sqrt{2} Y_{\rep2'',1}^{(1)} & 0 & \alpha_1
Y_{\rep2'',2}^{(1)} \\
0 & \alpha_1 \sqrt{2} Y_{\rep2'',2}^{(1)} & \alpha_1 Y_{\rep2'',1}^{(1)} \\
 \alpha_2 Y_{\rep{3},1}^{(2)} & \alpha_2 Y_{\rep{3},3}^{(2)} & \alpha_2 Y_{\rep{3},2}^{(2)} \\
\end{pmatrix}v_u
\end{aligned}$} } \\ 
$M_d$ & \multicolumn{10}{c}{{\small
$\begin{aligned}	\begin{pmatrix}
2 \beta_1 Y_{\rep{3},1}^{(2)} & \beta_2 Y_{\rep{3},3}^{(2)}-\beta_1 Y_{\rep{3},3}^{(2)} & -\beta_1 Y_{\rep{3},2}^{(2)}-\beta_2 Y_{\rep{3},2}^{(2)} \\
-\beta_1 Y_{\rep{3},3}^{(2)}-\beta_2 Y_{\rep{3},3}^{(2)} & 2 \beta_1 Y_{\rep{3},2}^{(2)} & \beta_2 Y_{\rep{3},1}^{(2)}-\beta_1 Y_{\rep{3},1}^{(2)} \\
\beta_2 Y_{\rep{3},2}^{(2)}-\beta_1 Y_{\rep{3},2}^{(2)} & -\beta_1 Y_{\rep{3},1}^{(2)}-\beta_2 Y_{\rep{3},1}^{(2)} & 2 \beta_1 Y_{\rep{3},3}^{(2)} \\
\end{pmatrix}v_d \end{aligned}$} } \\  \midrule
IIB& $u^c$ & $d^c$ & $Q$ & $H_{u/d}$ & $Y^{(2)}_{\rep{3}}$&
\multicolumn{1}{c}{}  &\multicolumn{1}{c}{} &\multicolumn{1}{c}{} &\multicolumn{1}{c}{} &  \\
$\rep{\rho}$ & $\rep2'' \oplus \rep{1}$ & $\rep{3}$ & $\rep{3}$ & $\rep{1}$ &  $\rep{3}$&
\multicolumn{1}{c}{}  &\multicolumn{1}{c}{} & \multicolumn{1}{c}{}&\multicolumn{1}{c}{} &  \\
$k$ & $(1,2)$ & $2$ & $1$ & $-1$ &$2$ &
 \multicolumn{1}{c}{} &\multicolumn{1}{c}{} &\multicolumn{1}{c}{} & \multicolumn{1}{c}{} & \\
\# of $Y$ & $-$ & $-$ & $-$ & $-$ & 1  &
\multicolumn{1}{c}{} &\multicolumn{1}{c}{} &\multicolumn{1}{c}{} & \multicolumn{1}{c}{} &  \\
$M_u$ & \multicolumn{10}{c}{{\small
$\begin{aligned}
\begin{pmatrix}
0 & 0 & 0 \\
0 & 0 & 0 \\
\alpha Y_{\rep{3},1}^{(2)} & \alpha Y_{\rep{3},3}^{(2)} & \alpha Y_{\rep{3},2}^{(2)} \\
\end{pmatrix}v_u
\end{aligned}$} } \\ 
$M_d$ & \multicolumn{10}{c}{{\small
$\begin{aligned}
\begin{pmatrix}
2 \beta_1 Y_{\rep{3},1}^{(2)} & \beta_2 Y_{\rep{3},3}^{(2)}-\beta_1 Y_{\rep{3},3}^{(2)} & -\beta_1 Y_{\rep{3},2}^{(2)}-\beta_2 Y_{\rep{3},2}^{(2)} \\
-\beta_1 Y_{\rep{3},3}^{(2)}-\beta_2 Y_{\rep{3},3}^{(2)} & 2 \beta_1 Y_{\rep{3},2}^{(2)} & \beta_2 Y_{\rep{3},1}^{(2)}-\beta_1 Y_{\rep{3},1}^{(2)} \\
\beta_2 Y_{\rep{3},2}^{(2)}-\beta_1 Y_{\rep{3},2}^{(2)} & -\beta_1 Y_{\rep{3},1}^{(2)}-\beta_2 Y_{\rep{3},1}^{(2)} & 2 \beta_1 Y_{\rep{3},3}^{(2)} \\
\end{pmatrix}v_d \end{aligned}$} } \\  \midrule
IIIB& $u^c$ & $d^c$ & $Q$ & $H_{u/d}$  & $Y^{(1)}_{\rep2'}$ & $Y^{(1)}_{\rep2''}$ & $Y^{(2)}_{\rep{3}}$  & $Y^{(2)}_{\rep{1}}$& $Y^{(2)}_{\rep{1''}}$ & \\ 
$\rep{\rho}$ & $\rep2'' \oplus \rep{1}$ & $\rep{3}$ & $\rep{3}$ & $\rep{1}$ & $\rep2'$ & $\rep2''$ & $\rep{3}$ & $\rep{1}$ & $\rep{1''}$ & \\ 
$k$ & $(1,2)$ & $2$ & $1$ & $-1$ & $3$ & $3$ & $2$ & $2$ & $2$ & \\ 
\# of $Y$ & $-$ & $-$ & $-$ & $-$ & 2 & 1 & 3 & 2 & 1 &  \\ 
$M_u$ & \multicolumn{10}{c}{{\small
$\begin{aligned}
\begin{pmatrix}
-\sqrt{2} \mathcal{Y}_{\rep2'',1}^{(1)}-\mathcal{Y}_{\rep2',2}^{(1)} & \sqrt{2} \mathcal{Y}_{\rep2',1}^{(1)} & \mathcal{Y}_{\rep2'',2}^{(1)} \\
-\mathcal{Y}_{\rep2',1}^{(1)} & \sqrt{2} \mathcal{Y}_{\rep2'',2}^{(1)} & \mathcal{Y}_{\rep2'',1}^{(1)}-\sqrt{2} \mathcal{Y}_{\rep2',2}^{(1)} \\
\sqrt{2} \mathcal{Y}_{\rep{3},1}^{(2)} & \sqrt{2} \mathcal{Y}_{\rep{3},3}^{(2)} & \sqrt{2} \mathcal{Y}_{\rep{3},2}^{(2)} \\
\end{pmatrix}
v_u
\end{aligned}$} } \\ 
$M_d$ & \multicolumn{10}{c}{{\small
$\begin{aligned}\begin{pmatrix}
			\sqrt{2}\mathcal{Y}_{\rep{1}}^{(2)}+2 \mathcal{Y}_{\rep{3},1}^{(2)} & \sqrt{2} \mathcal{Y}_{\rep{1}''}^{(2)} - \mathcal{Y}_{\rep{3},3}^{(2)} + \mathcal{X}_{\rep{3},3}^{(2)}  & -\mathcal{Y}_{\rep{3},2}^{(2)} -  \mathcal{X}_{\rep{3},2}^{(2)} \\
			\sqrt{2} \mathcal{Y}_{\rep{1}''}^{(2)} - \mathcal{Y}_{\rep{3},3}^{(2)} -  \mathcal{X}_{\rep{3},3}^{(2)} & 2 \mathcal{Y}_{\rep{3},2}^{(2)} & \sqrt{2} \mathcal{Y}_{\rep{1}}^{(2)} -\mathcal{Y}_{\rep{3},1}^{(2)} + \mathcal{X}_{\rep{3},1}^{(2)} \\
			-\mathcal{Y}_{\rep{3},2}^{(2)} + \mathcal{X}_{\rep{3},2}^{(2)}  & \sqrt{2} \mathcal{Y}_{\rep{1}}^{(2)} - \mathcal{Y}_{\rep{3},1}^{(2)} -  \mathcal{X}_{\rep{3},1}^{(2)} & \sqrt{2} \mathcal{Y}_{\rep{1}''}^{(2)} + 2 \mathcal{Y}_{\rep{3},3}^{(2)} \\
		   \end{pmatrix}v_d \end{aligned}$} } \\ \bottomrule
\end{tabular}
\caption{\label{tab:ModelsummaryB}
Summary of features of modular flavor models, where the up-quark modular weights are $k_{u^c_{1,2,3}}=1,1,2$ and all other charges are given in \Cref{eq:Model32RepWeight}. 
The determinants of the mass matrices $M_{u}$ turn out to be both zero for the models IB and IIB and that of $M_{d}$ is the same as in \Cref{eq:determinantMd_allModels}.}
\end{table}

In order to explore the effects of a different choice of modular weights, let us perform a very minor variation of our models defined in \Cref{eq:Model32RepWeight}. 
Let us only consider that  $k_{u^c_{1,2,3}}=1,1,2$ instead, i.e.\ $u^c_D$ has modular weight 1 instead of 3. 
This choice alters the renormalizable superpotential, as different weights of modular forms are now required to fulfill modular invariance. 
Repeating the method described in \Cref{sec:ModelIA,sec:ModelIIA,sec:ModelIIIA} for this new choice, we find the results listed in \Cref{tab:ModelsummaryB}. 
Model IB is based on the group $\Gamma'_3$, model IIB on $2T$, and model IIIB on $\Gamma(2)/\Gamma(6)$.
Observe that all $M_u$ matrices differ, now leading to more zeroes than in the previous case. 
In particular the structure of the mass matrix in the model IIB disables the possibility to arrive at three nonvanishing up-quark masses. 
As discussed in \Cref{app:fittingResults}, only model IIIB is best equipped to reproduce observations.

Despite the strong differences among the various realizations of the modular $T'$ flavor symmetry, let us briefly discuss the universal characteristics and potential connections that exist in the models:
\begin{itemize}
\item[1.] Firstly, due to their identical finite group structure, some of the mass textures are actually identical, as shown in \Cref{tab:ModelsummaryA,tab:ModelsummaryB} for some choice of matter representations. This applies particularly to the structure of $M_u$, which arises from choosing a triplet $T'$ representation, which is universal for all realizations of $T'$ discussed here. Therefore, we can expect that the modular symmetry origin of texture zeros discussed in previous literature~\cite{Lu:2019vgm,Zhang:2019ngf,Ding:2022aoe,Nomura:2023usj} remains valid for all three realizations of $T'$, but the specific implementation conditions will be different.

\item[2.] Secondly, there are hidden relations among some mass-matrix entries of different realizations of $T'$. For instance, the triplet modular forms $Y^{(2)}_{\rep3}$ of weight 2 coincide with those of model IA, IIA and IIIA; the doublet modular forms $Y^{(1)}_{\rep2''}$ of weight 1 in model IB and IIIB are exactly same; and there are some algebraic relations between the doublet modular forms in model IIA and IB, as shown in Appendix~\ref{app:relations}.
Thus, in principle, some relations among these models may be relevant. It seems an inviting challenge to try and express these relations explicitly, possibly via the study of modular invariants~\cite{Chen:2024otk}. This question will be explored further elsewhere.

\item[3.] Thirdly, regarding the near-critical behavior of fermion masses, each of the models has a different scaling behavior (cf.\ e.g.\ \Cref{eq:massHierarchyIA,eq:massHierarchyIIA}), depending on the specific form of the representation matrices. However, the determinants of the mass matrices (i.e.\ products of quark masses) are predicted both in model IA and model IIA to be proportional to $\eta^{16}(\tau)$, see \Cref{eq:determinantMu_IA,eq:determinantMu_IIA}.
This can be attributed to the fact that the determinant of the mass matrix is a 1-dimensional \ac{VVMF}, derived solely from the product of determinants of matter field representations. For models IA and IIA, this modular form builds a $\rep{1}''$ representation.
\end{itemize}

\section{Conclusions}
\label{sec:conclusion}
	
Model building based on modular flavor symmetries usually involves four steps:
\begin{enumerate}
 \item Select a particular finite modular group, typically one from the set of $\SL{2,\mathds{Z}}/\Gamma(N)$ groups.
 \item Assign the representations and modular weights of the matter fields.
 \item Find the set of \acp{VVMF} associated with the chosen finite modular group.
 \item Scan the parameter space to identify the point that best fit the model to data.
\end{enumerate}

In this work, we have shown that this procedure has more variations than previously appreciated.
In general, there are multiple modular flavor symmetries which lead to the same finite modular group. 
This gives rise to different sets of \acp{VVMF}, and the resulting phenomenological properties differ in critical aspects.

Focusing on $T'$ as a benchmark finite modular symmetry, we have studied three sample models based on this finite modular group.
In these models, $T'$ originates from the quotients $\Gamma'_3=\SL{2,\mathds{Z}}/\Gamma(3)$, $2T=\SL{2,\mathds{Z}}/N_{[24,3]II}$ and $\Gamma(2)/\Gamma(6)$, respectively.
We have shown that, depending on how $T'$ is realized,
the finite modular group can act differently on the modulus $\tau$ and matter fields.
Consequently, also the corresponding \acp{VVMF} differ in many important aspects: in a given representation and at a given weight, different amounts of modular forms may exist, and even if there is a unique modular for a given representation and modular weight, they can be different functions.

To illustrate our findings, we constructed quark models based on these three different realizations of $T'$ with identical representation and weight assignments. 
We found that each model can have a different number of parameters, different mass textures, and different near-critical behaviors. 
If a given realization is unable to fit data, another realization of the same finite modular group can be compatible with observations.
In particular, the ability of a given modular flavor symmetry to accommodate hierarchies depends very sensitively on the realization.
According to what we find, it seems much more straightforward to explain hierarchies via the properties of modular forms if one goes beyond the usual congruence subgroups of $\SL{2,\mathds{Z}}$.
This means that disfavoring, or ruling out, a given finite modular group requires a far more extensive analysis than previously appreciated.
We have also uncovered some (algebraic) relations among the \acp{VVMF} of the various instances of the finite modular group, which may hint at some subtle relations among the models. 
The detailed exploration of these relations is left for future works.

Our analysis was conducted in a bottom-up approach. This allowed us to dial representations and modular weights at will.
The top-down approach is far more restrictive in these regards.
Therefore, it will be interesting to find out to which extent the realizations of finite modular symmetries have \ac{UV} completions.

\clearpage

\section*{Acknowledgments}
This work was supported by UC-MEXUS-CONACyT grant No.\ CN-20-38.
CAO, RDC and SRS are supported by UNAM-PAPIIT IN113223, and Marcos Moshinsky Foundation.
MCC, XGL and MR are supported by U.S.\ National Science Foundation under Grant No.~PHY-2210283.
We are thankful to the Mainz Institute for Theoretical Physics (MITP) of the Cluster of Excellence
PRISMA+ (Project ID 390831469), for its hospitality and partial support at some stage of this work.
SRS acknowledges UC-Irvine for the welcoming atmosphere offered during parts of this work.
	
	
\begin{appendix}

\section{Tensor product rules for \texorpdfstring{$T'$}{T(prime)}}
\label{app:GroupTprime}

Some representations of the discrete group $T'\cong[24,3]$ studied in this paper have been presented in \Cref{tab:Tprime_reps,tab:Quotient_reps}. In addition, \Cref{tab:Tprime_character} corresponds to the character table of $T'$ and displays representatives of each conjugacy class for three instances of $T'$.

\begin{table}[h!]
	\centering
		\renewcommand{\tabcolsep}{2.8mm}
		\renewcommand{\arraystretch}{1.3}
		\begin{tabular}{ccccccccccccccc}\toprule
			\text{Classes}  &  $1C_{1}$  & $1C_{2}$ &  $4C_{3}$   &  $4C'_{3}$   &  $6C_{4}$   &  $4C_{6}$   &  $4C'_{6}$ \\ 
			\midrule
			\text{Representative for $\Gamma'_{3}$}          &  $1$   & $S^2$ &  $T^{2}$  &  $T$     &  $S$  &  $TS^2$    & $ST^{2}$  \\  
			\midrule
			\text{Representative for $2T$}                   &  $1$   & $S^2$ &  $T^{2}$  &  $T^{4}$ &  $S$  &  $T$       & $T^{5}$  \\  
			\midrule
			\text{Representative for $\Gamma(2)/\Gamma(6)$}  &  $1$   & $a^2$ &  $b^{2}$  &  $b$     &  $a$  &  $ba^2$    & $ab^{2}$  \\  
			\midrule
			$\rep{1}$    &  $1$  &  $1$  &  $1$        &  $1$         &  $1$  &  $1$        &  $1$  \\ 
			$\rep{1}'$   &  $1$  &  $1$  &  $\omega^2$ &  $\omega$    &  $1$  &  $\omega$   &  $\omega^2$  \\ 
			$\rep{1}''$  &  $1$  &  $1$  &  $\omega $  &  $\omega^2$  &  $1$  &  $\omega^2$ &  $\omega $  \\ 
			$\rep{2}$    &  $2$  &  $-2$ &  $-1$       &  $-1$        &  $0$  &  $1$        &  $1$   \\ 
			$\rep{2}'$   &  $2$  &  $-2$ &  $-\omega^2$&  $-\omega $  &  $0$  &  $\omega $  &  $\omega^2$   \\ 
			$\rep{2}''$  &  $2$  &  $-2$ &  $-\omega$  &  $-\omega^2$ &  $0$  &  $\omega^2$ &  $\omega $   \\ 
			$\rep{3}$    &  $3$  &  $3$  &  $0$        &  $0$         & $-1$  &  $0$        &  $0$  \\ \bottomrule
		\end{tabular}
		\caption{\label{tab:Tprime_character}%
		Character table of the discrete group $T'\cong[24,3]$ with $\omega=\ee^{2\pi \ii/3}$. Here, $S$ and $T$ are the
		standard \SL{2,\mathds{Z}} generators, $a=T\, S\,T^4\, S^3\, T$, and $b=T^4$.}
\end{table}
The tensor product decomposition and \ac{CG} coefficients of $T'$ (in the basis chosen in this work) are given by
\begin{align}
	\left( x_1 \right)_{\rep1'} \otimes \left( y_1 \right)_{\rep1'}&=\left( x_1 y_1 \right)_{\rep1''}\;,\notag\\ 
	\left( x_1 \right)_{\rep1'} \otimes \left( y_1 \right)_{\rep{1}''}&=\left( x_1 y_1 \right)_{\rep{1}}\;, \notag\\ 
	\left( x_1 \right)_{\rep1''} \otimes \left( y_1 \right)_{\rep1''}&=\left( x_1 y_1 \right)_{\rep1'}\;,\notag\\
	\begin{pmatrix} x_{1} \end{pmatrix}_{\rep{1}} \otimes \begin{pmatrix} y_{1}\\ y_{2}  \end{pmatrix}_{\rep{2}}&=\begin{pmatrix} x_{1} \end{pmatrix}_{\rep1'} \otimes \begin{pmatrix} y_{1}\\ y_{2}  \end{pmatrix}_{\rep2''}=\begin{pmatrix} x_{1} \end{pmatrix}_{\rep1''} \otimes \begin{pmatrix} y_{1}\\ y_{2}  \end{pmatrix}_{\rep2'}=x_{1}\begin{pmatrix} y_{1} \\ y_{2} \end{pmatrix}_{\rep{2}} \;,\notag\\
	\begin{pmatrix} x_{1} \end{pmatrix}_{\rep{1}} \otimes \begin{pmatrix} y_{1}\\ y_{2}  \end{pmatrix}_{\rep2'}&=\begin{pmatrix} x_{1} \end{pmatrix}_{\rep1'} \otimes \begin{pmatrix} y_{1}\\ y_{2}  \end{pmatrix}_{\rep{2}}=\begin{pmatrix} x_{1} \end{pmatrix}_{\rep1''} \otimes \begin{pmatrix} y_{1}\\ y_{2}  \end{pmatrix}_{\rep2''}=x_{1}\begin{pmatrix} y_{1} \\ y_{2} \end{pmatrix}_{\rep2'} \;,\notag\\
	\begin{pmatrix} x_{1} \end{pmatrix}_{\rep{1}} \otimes \begin{pmatrix} y_{1}\\ y_{2}  \end{pmatrix}_{\rep2''}&=\begin{pmatrix} x_{1} \end{pmatrix}_{\rep1'} \otimes \begin{pmatrix} y_{1}\\ y_{2}  \end{pmatrix}_{\rep2'}=\begin{pmatrix} x_{1} \end{pmatrix}_{\rep1''} \otimes \begin{pmatrix} y_{1}\\ y_{2}  \end{pmatrix}_{\rep{2}}=x_{1}\begin{pmatrix} y_{1} \\ y_{2} \end{pmatrix}_{\rep2''} \;,\notag\\
	\begin{pmatrix} x_{1} \end{pmatrix}_{\rep{1}} \otimes \begin{pmatrix} y_{1}\\ y_{2} \\ y_{3} \end{pmatrix}_{\rep{3}}&=x_{1}\begin{pmatrix} y_{1} \\ y_{2} \\ y_{3} \end{pmatrix}_{\rep{3}} \;,\notag\\
	\begin{pmatrix} x_{1} \end{pmatrix}_{\rep1'} \otimes \begin{pmatrix} y_{1}\\ y_{2} \\ y_{3} \end{pmatrix}_{\rep{3}}&=x_{1}\begin{pmatrix} y_{3} \\ y_{1} \\ y_{2} \end{pmatrix}_{\rep{3}} \;,\notag\\
	\begin{pmatrix} x_{1} \end{pmatrix}_{\rep1''} \otimes \begin{pmatrix} y_{1}\\ y_{2} \\ y_{3} \end{pmatrix}_{\rep{3}}&=x_{1}\begin{pmatrix} y_{2} \\ y_{3} \\ y_{1} \end{pmatrix}_{\rep{3}} \;,\notag\\
	\begin{pmatrix} x_{1}\\ x_{2} \end{pmatrix}_{\rep{2}} \otimes \begin{pmatrix} y_{1}\\ y_{2} \end{pmatrix}_{\rep{2} }&=\begin{pmatrix} x_{1}\\ x_{2} \end{pmatrix}_{\rep2'} \otimes \begin{pmatrix} y_{1}\\ y_{2} \end{pmatrix}_{\rep2''}=\left(x_{1}y_{2} - x_{2}y_{1}\right)_{\rep{1}} \oplus \begin{pmatrix} x_{1}y_{2} + x_{2}y_{1}\\ \sqrt{2}x_{2}y_{2}\\ -\sqrt{2}x_{1}y_{1} \end{pmatrix}_{\rep{3}}\;,\notag\\
	\begin{pmatrix} x_{1}\\ x_{2} \end{pmatrix}_{\rep2' } \otimes \begin{pmatrix} y_{1}\\ y_{2} \end{pmatrix}_{\rep2'}&=\begin{pmatrix} x_{1}\\ x_{2} \end{pmatrix}_{\rep{2}} \otimes \begin{pmatrix} y_{1}\\ y_{2} \end{pmatrix}_{\rep2''}=\left(x_{1}y_{2} - x_{2}y_{1}\right)_{\rep1''} \oplus \begin{pmatrix} \sqrt{2}x_{2}y_{2}\\ -\sqrt{2}x_{1}y_{1} \\ x_{1}y_{2} + x_{2}y_{1}  \end{pmatrix}_{\rep{3}}\;,\notag\\
	\begin{pmatrix} x_{1}\\ x_{2} \end{pmatrix}_{\rep2''} \otimes \begin{pmatrix} y_{1}\\ y_{2} \end{pmatrix}_{\rep2''}&=\begin{pmatrix} x_{1}\\ x_{2} \end{pmatrix}_{\rep{2}} \otimes \begin{pmatrix} y_{1}\\ y_{2} \end{pmatrix}_{\rep2'}=\left(x_{1}y_{2} - x_{2}y_{1}\right)_{\rep1'} \oplus \begin{pmatrix} -\sqrt{2}x_{1}y_{1} \\ x_{1}y_{2} + x_{2}y_{1}\\ \sqrt{2}x_{2}y_{2} \end{pmatrix}_{\rep{3}}\;,\notag\\
	\begin{pmatrix} x_{1}\\ x_{2} \end{pmatrix}_{\rep{2}} \otimes \begin{pmatrix} y_{1}\\ y_{2} \\ y_{3} \end{pmatrix}_{\rep{3}}&=\begin{pmatrix} x_{1} y_{1}+ \sqrt{2}x_{2} y_{3} \\ \sqrt{2} x_{1} y_{2} -x_{2} y_{1} \end{pmatrix}_{\rep{2}} \oplus \begin{pmatrix} x_{1} y_{2}+ \sqrt{2}x_{2} y_{1} \\ \sqrt{2}x_{1} y_{3} -x_{2} y_{2} \end{pmatrix}_{\rep2'} \oplus \begin{pmatrix} x_{1} y_{3}+ \sqrt{2}x_{2} y_{2} \\ \sqrt{2}x_{1} y_{1} -x_{2} y_{3} \end{pmatrix}_{\rep2''}\;,\notag\\
	\begin{pmatrix} x_{1}\\ x_{2} \end{pmatrix}_{\rep2'} \otimes \begin{pmatrix} y_{1}\\ y_{2} \\ y_{3} \end{pmatrix}_{\rep{3}}&=\begin{pmatrix} x_{1} y_{3}+ \sqrt{2}x_{2} y_{2} \\ \sqrt{2}x_{1} y_{1} -x_{2} y_{3} \end{pmatrix}_{\rep{2}} \oplus \begin{pmatrix} x_{1} y_{1}+ \sqrt{2}x_{2} y_{3} \\ \sqrt{2}x_{1} y_{2} -x_{2} y_{1} \end{pmatrix}_{\rep2'} \oplus \begin{pmatrix} x_{1} y_{2}+ \sqrt{2}x_{2} y_{1} \\ \sqrt{2}x_{1} y_{3} -x_{2} y_{2} \end{pmatrix}_{\rep2''}\;,\notag\\
	\begin{pmatrix} x_{1}\\ x_{2} \end{pmatrix}_{\rep2''} \otimes \begin{pmatrix} y_{1}\\ y_{2} \\ y_{3} \end{pmatrix}_{\rep{3}}&=\begin{pmatrix} x_{1} y_{2}+ \sqrt{2}x_{2} y_{1} \\ \sqrt{2}x_{1} y_{3} -x_{2} y_{2} \end{pmatrix}_{\rep{2}} \oplus \begin{pmatrix} x_{1} y_{3}+ \sqrt{2}x_{2} y_{2} \\ \sqrt{2}x_{1} y_{1} -x_{2} y_{3} \end{pmatrix}_{\rep2'} \oplus \begin{pmatrix} x_{1} y_{1}+ \sqrt{2}x_{2} y_{3} \\ \sqrt{2}x_{1} y_{2} -x_{2} y_{1} \end{pmatrix}_{\rep2''}\;,\notag\\
	\begin{pmatrix} x_{1}\\ x_{2} \\ x_{3} \end{pmatrix}_{\rep{3}} \otimes \begin{pmatrix} y_{1}\\ y_{2} \\ y_{3} \end{pmatrix}_{\rep{3}}&=\begin{pmatrix} x_{1} y_{1} + x_{2} y_{3} + x_{3} y_{2}  \end{pmatrix}_{\rep{1}} \oplus \begin{pmatrix} x_{1} y_{2} + x_{2} y_{1} + x_{3} y_{3}  \end{pmatrix}_{\rep1'} \oplus \begin{pmatrix} x_{1} y_{3} + x_{2} y_{2} + x_{3} y_{1}  \end{pmatrix}_{\rep1''}\notag\\
	&\qquad{}\oplus
	\begin{pmatrix} 2 x_{1} y_{1} - x_{2} y_{3} - x_{3} y_{2} \\ 2 x_{3} y_{3} - x_{1} y_{2} - x_{2} y_{1}  \\ 2 x_{2} y_{2}- x_{1} y_{3}  - x_{3} y_{1} \end{pmatrix}_{\rep{3}_\mathrm{S}} \oplus \begin{pmatrix} x_{3} y_{2}- x_{2} y_{3}  \\  x_{2} y_{1}- x_{1} y_{2}  \\ x_{1} y_{3} - x_{3} y_{1} \end{pmatrix}_{\rep{3}_\mathrm{A}}\;.\label{eq:Tprime_tensorProducts}
\end{align}
Here, the subscripts ``S'' and ``A'' denote symmetric and antisymmetric contractions, respectively.

\section{Algebraic relations of modular form multiplets of \texorpdfstring{$2T$}{2T} and \texorpdfstring{$\Gamma'_3$}{Gamma(prime,3)}}
\label{app:relations}

All even-weight \acp{VVMF} of $\Gamma'_3$ and $2T$ are identical and can be constructed from the tensor products of their respective lowest weight modular form doublets. This leads to an algebraic relation between the modular form doublets of the two different symmetries $\Gamma'_3$ and $2T$. It is not difficult to see that this algebraic relation between $Y^{(1)}_{(\Gamma'_3)\rep{2''}}$ and $Y^{(3)}_{(2T)\rep{2'}}$ is given by
\begin{equation}
	\left(Y^{(1)}_{(\Gamma'_3)\rep{2''}}\right)^{\otimes 6}_{\rep{3}} \eqdef Y^{(6)}_{(\Gamma'_3)\rep{3}} =
	Y^{(6)}_{(2T)\rep{3}} \defeq \left(Y^{(3)}_{(2T)\rep{2'}}\right)^{\otimes 2}_{\rep{3}}  \;.
\end{equation}
The specific tensor product leads to
\begin{equation}
	\label{eq:RelationsContraction}
	\left(Y^{(1)}_{(\Gamma'_3)\rep{2''}}\right)^{\otimes 6}_{\rep{3}}=\begin{pmatrix}
		-\frac{1}{12}X_2^3(\sqrt{2}X_2^3-4X_1^3) \\ \frac{1}{12}(-4X_1^5X_2+\sqrt{2}X_1^2X_2^4)\\ \frac{1}{6}X_1X_2^2(2\sqrt{2}X_1^3-X_2^3)
	\end{pmatrix}=\begin{pmatrix}
		\sqrt{2}Y_2^2 \\ -\sqrt{2}Y_1^2 \\ 2Y_1Y_2
	\end{pmatrix}=\left(Y^{(3)}_{(2T)\rep{2'}}\right)^{\otimes 2}_{\rep{3}}\;,
\end{equation}
where $X_{1,2}$ and $Y_{1,2}$ are the components of $Y^{(1)}_{(\Gamma'_3)\rep{2''}}$ and $Y^{(3)}_{(2T)\rep{2'}}$, respectively.
\Cref{eq:RelationsContraction} yields the algebraic relations
\begin{subequations}\label{eq:indenpendtReations}
\begin{align}	
		Y_2^2&=\frac{\sqrt{2}}{6} X_1^3 \,X_2^3 -\frac{1}{12} X_2^6\;,\\
		Y_1^2&=\frac{\sqrt{2}}{6} X_1^5\, X_2 -\frac{1}{12} X_1^2\,X_2^4\;,\\
		Y_1\,Y_2&=\frac{\sqrt{2}}{6} X_1^4 X_2^2 -\frac{1}{12} X_1\,X_2^5\;.
\end{align}
\end{subequations}
Since other higher weight \acp{VVMF} can be constructed by tensor products of these lowest weight modular forms, the independent algebraic relations between $2T$ and $\Gamma_3'$ can be deduced by~\Cref{eq:indenpendtReations}. We have verified that there are no additional nontrivial algebraic relations.

\section{Properties of modular forms for \texorpdfstring{$\Gamma(2) / \Gamma(6)$}{Gamma(2)/Gamma(3)}}
\label{app:table_expansions}

All the $q$-expansions for the different \acp{VVMF} of weights 2 and 3 for $\Gamma(2)/\Gamma(6)$, listed in~\Cref{eq:Contractions,eq:weight3G2G6}, are presented in~\Cref{tab:Gamma26Expansions}.

As we mentioned in~\Cref{sec:Tprime}, not all the \acp{VVMF} of weight 1 of $\Gamma(6)$ in \Cref{eq:weight1FormsG6} are algebraically independent. Following a similar procedure as in~\Cref{app:relations}, it is possible to derive a complete set of nontrivial algebraic relations among the scalar weight-1 modular forms presented in~\Cref{eq:weight1FormsG6}. 
These algebraic relations are
\begin{align}
		Y_4 \,Y_6 - \tfrac{1}{2} (Y_1 Y_6 + Y_2 Y_5)&=0\;, \qquad&	\sqrt{2} \, Y_3\, Y_5 + Y_2\, Y_6&=0\;,\notag\\
		Y_1\, Y_5 + Y_3\, Y_6 + Y_4\, Y_5&=0\;,               	  &	-Y_1^2 + Y_6^2 + Y_4^2&=0\;, \notag\\
		-\sqrt{2} \, Y_1\, Y_2 + Y_3^2 + Y_5^2 &=0\;,  		  &	Y_2^2 + \sqrt{2} \, (Y_3\, Y_4 + Y_5\, Y_6)&=0\;, \notag\\
		\sqrt{2} \, Y_2\, Y_4 - Y_3^2 + Y_5^2&=0\;, 			  &	Y_1\, Y_4 + Y_2\, Y_3 + Y_4^2 - Y_6^2&=0\;, \notag\\
		Y_1\, Y_3 + Y_5\, Y_6 - Y_3\, Y_4&=0\;.\label{eq:G6relations}
\end{align}

\begin{table}[t!]
 \centering
 \begin{tabular}{cc}
	\toprule
	Modular form & $q$-expansion \\
	\midrule
	$Y^{(2)}_{\rep{1}I}$ & $1 + 24 q + 24 q^2 + 96 q^3 + 24 q^4 + 144 q^5 + 96 q^6 + 192 q^7 + 24 q^8 +\cdots$ \\
			$Y^{(2)}_{\rep{1}II}$ & $-8\sqrt{3}q^{1/2}\left(1 + 4 q + 6 q^2 + 8 q^3 + 13 q^4 + 12 q^5 + 14 q^6 + 24 q^7 +\cdots\right)$ \\
			$Y^{(2)}_{\rep{1''}}$ & $1 - 4 q + 2 q^2 + 8 q^3 - 5 q^4 - 4 q^5 - 10 q^6 + 8 q^7 + 9 q^8+\cdots$ \\
			$Y^{(2)}_{\rep{3}I}$ & $\begin{pmatrix}   -1 + 12 q + 12 q^2 + 12 q^3 + 12 q^4 + 72 q^5 + 12 q^6 + 96 q^7 + 12 q^8  + \cdots \\
 					-6q^{1/3}\left( 1 + q + 8 q^2 + 6 q^3 + 14 q^4 + q^5 + 20 q^6 + 12 q^7 + \cdots \right) \\
					-6 q^{2/3}\left( 1 + 6 q + q^2 + 12 q^3 + 8 q^4 + 18 q^5 + 6 q^6 + 24 q^7  + \cdots \right)
				   \end{pmatrix}$ \\
$Y^{(2)}_{\rep{3}II}$ & $\begin{pmatrix} -4\sqrt{3}q^{1/2}\left(1 + q + 6 q^2 + 8 q^3 + q^4 + 12 q^5 + 14 q^6 + 6 q^7 +\cdots \right)\\											12\sqrt{3}q^{5/6}\left(1 + 2 q + 3 q^2 + 4 q^3 + 5 q^4 + 8 q^5 + 7 q^6 + 8 q^7 +\cdots\right)\\
					2\sqrt{3}q^{1/6}\left(1 + 8 q + 14 q^2 + 20 q^3 + 31 q^4 + 32 q^5 + 38 q^6 + 44 q^7 +\cdots\right)
					 \end{pmatrix}$ \\
$Y^{(2)}_{\rep{3}III}$ & $\begin{pmatrix} -\sqrt{2}\left(1 + 12 q + 36 q^2 + 12 q^3 + 84 q^4 + 72 q^5 + 36 q^6 + 96 q^7 +
 					180 q^8 +\cdots\right)\\
					6\sqrt{2}q^{1/3}\left(1 + 7 q + 8 q^2 + 18 q^3 + 14 q^4 + 31 q^5 + 20 q^6 + 36 q^7 +\cdots\right)\\
					18\sqrt{2}q^{2/3}\left(1 + 2 q + 5 q^2 + 4 q^3 + 8 q^4 + 6 q^5 + 14 q^6 + 8 q^7 +\cdots\right)
					\end{pmatrix}$ \\
$Y^{(3)}_{\rep{2}I}$ & $\begin{pmatrix} -12q^{1/3}\left(1 - 2 q - 4 q^2 + 6 q^3 + 8 q^4 + 4 q^5 - 16 q^6 - 24 q^7 +\cdots\right)\\
 					12\sqrt{2}q^{2/3}\left(1 - 3 q - 2 q^2 + 12 q^3 - 4 q^4 - 9 q^5 + 6 q^6 - 12 q^7+\cdots\right)
					 \end{pmatrix}$\\
$Y^{(3)}_{\rep{2}II}$ & $\begin{pmatrix} -12\sqrt{3}\left(1 - 4 q + 3 q^2 + 4 q^3 - q^4 - 4 q^5 - 11 q^6 + 20 q^7 - 9 q^8+\cdots\right)\\
 					2\sqrt{6}q^{1/6}\left(1 - 4 q + 8 q^2 - 16 q^3 + 7 q^4 + 44 q^5 - 34 q^6 - 40 q^7 +\cdots\right)
					\end{pmatrix}$\\
$Y^{(3)}_{\rep{2'}I}$ & $\begin{pmatrix} -6q^{2/3}\left(7 + 24 q + 103 q^2 + 120 q^3 + 350 q^4 + 288 q^5 + 600 q^6 +
 					528 q^7  +\cdots\right) \\
 					-\sqrt{2}\left(1 - 24 q - 210 q^2 - 240 q^3 - 600 q^4 - 720 q^5 - 1722 q^6 -
 					1200 q^7 - 3090 q^8+\cdots\right)
					\end{pmatrix}$\\
$Y^{(3)}_{\rep{2'}II}$ & $\begin{pmatrix} 2\sqrt{3}\left(1 + 50 q + 170 q^2 + 362 q^3 + 601 q^4 + 962 q^5 + 1370 q^6 +
 					1850 q^7 + 2451 q^8 +\cdots\right)\\
					-2\sqrt{6}\left(5 + 41 q + 96 q^2 + 250 q^3 + 365 q^4 + 480 q^5 + 850 q^6 + 960 q^7 +
					1152 q^8 +\cdots\right)
					\end{pmatrix}$\\
$Y^{(3)}_{\rep{2'}III}$ & $\begin{pmatrix} 54\sqrt{2}\left(1 + 8 q + 17 q^2 + 40 q^3 + 50 q^4 + 96 q^5 + 104 q^6 + 176 q^7 +
					 170 q^8 +\cdots\right)\\
					 -2\left(1 + 72 q + 270 q^2 + 720 q^3 + 936 q^4 + 2160 q^5 + 2214 q^6 +
 					3600 q^7 + 4590 q^8 +\cdots\right)
					\end{pmatrix}$\\
$Y^{(3)}_{\rep{2'}IV}$ & $\begin{pmatrix} -\sqrt{6}q^{1/6}\left(1 + 2 q - 22 q^2 + 26 q^3 + 25 q^4 - 46 q^5 + 26 q^6 - 22 q^7 +\cdots\right)\\
					 -6\sqrt{3}q^{1/2}\left(1 - 3 q + 2 q^3 + 9 q^4 - 22 q^6 +\cdots\right)
					\end{pmatrix}$\\
$Y^{(3)}_{\rep{2''}I}$ & $\begin{pmatrix} 1 + 30 q + 168 q^2 + 246 q^3 + 750 q^4 + 576 q^5 + 1680 q^6 +
 					1500 q^7 + 2472 q^8 +\cdots\\
					3\sqrt{2}q^{1/3}\left(1 + 25 q + 50 q^2 + 168 q^3 + 170 q^4 + 409 q^5 + 362 q^6 + 840 q^7 +\cdots\right)
					\end{pmatrix}$\\
$Y^{(3)}_{\rep{2''}II}$ & $\begin{pmatrix} -8\sqrt{3}q^{1/2}\left(1 + 10 q + 30 q^2 + 50 q^3 + 91 q^4 + 150 q^5 + 170 q^6 + 246 q^7 +\cdots\right)\\
					-24\sqrt{6}\left(1 + 5 q + 12 q^2 + 22 q^3 + 35 q^4 + 50 q^5 + 70 q^6 + 92 q^7 +
 117 q^8+\cdots\right)
					\end{pmatrix}$\\
$Y^{(3)}_{\rep{2''}III}$ & $\begin{pmatrix}  -\left(1 + 30 q + 168 q^2 + 246 q^3 + 750 q^4 + 576 q^5 + 1680 q^6 +
 1500 q^7 + 2472 q^8 +\cdots\right)\\
  -3\sqrt{2}\left(1 + 25 q + 50 q^2 + 168 q^3 + 170 q^4 + 409 q^5 + 362 q^6 + 840 q^7 + 601 q^8+\cdots\right)
					\end{pmatrix}$\\
	\bottomrule
\end{tabular}
\caption{\label{tab:Gamma26Expansions}Explicit $q$-expansions of modular forms with weights 2 and 3 of the group $\Gamma(2)/\Gamma(6)$.}
\end{table}

\section{Fitting parameters of models with modular \texorpdfstring{$T'$}{T(prime)} symmetries}
\label{app:fittingResults}

Let us now summarize the results of a scan of the parameter space of models presented in \Cref{sec:models}, based on three instances of the $T'$ modular flavor symmetry.
The aim of our scan is to get a rough picture, and we thus ignore quantum corrections \cite{Luo:2002ti} as well as non-minimal kinetic terms \cite{Chen:2019ewa}.
The data we use for comparing our predictions is given in \cite[Table~3]{Ding:2023ydy}, where we consider the three mixing angles and the \CP-violation phase of the \ac{CKM} matrix as well as the two independent mass ratios in the up- and down-quark sectors.
For each of the scanned points, we compute the value of a global $\chi^2$ function, defined as usual by
\begin{equation}
  \chi^2 \defeq \sum_a \frac{(x^\text{pred}_a - x^\text{obs}_a)^2}{\sigma_a^2}\;,
\end{equation}
where the sum runs over the quark observables in consideration, $x^\text{obs}_a$ and $x^\text{pred}_a$ denote the $a$\textsuperscript{th} observable as obtained, respectively, from~\cite{Ding:2023ydy} and by direct computation in our models, and $\sigma_a$ is the experimental error associated with $x^\text{obs}_a$. 
We then identify the parameter points $\{\tau,\alpha_i,\beta_j\}$ that minimize $\chi^2$ and take them as the prediction of our models. 
In what follows, we list our results for three realizations of $T'$ studied in this work with two slightly different choices of modular weights and identical $T'$ representations.

We start with the models whose defining representations and modular weights are given in \Cref{eq:Model32RepWeight}, i.e.
\begin{subequations}
\begin{align}
	\rho^{(u^c)} &: \rep{2}'' \oplus \rep{1}\;, \quad & k_{u^c_{1,2,3}} &= (3,3,2)\;,\quad &\rho^{(d^c)}     &: \rep{3}\;, \quad  & k_{d^c}     &= 2\;, \\
	\rho^{(Q)}   &: \rep{3}\;,                      & k_{Q}           &= 1\;,         &\rho^{(H_{u/d})} &: \rep{1}\;,      & k_{H_{u/d}} &= -1\;.
\end{align}
\end{subequations}
For model IA with $\Gamma'_3$ finite modular group, the small number of parameters of the model makes it challenging to fit the model to data, as expected from the near-critical analysis discussed in \Cref{eq:massHierarchyIA,eq:massHierarchyIA2}. 
At the minimum of $\chi^2$, we obtain the best-fit values of observables
\begin{align}
	\sin^2 \theta_{12} &= 0.0879\;, & \sin^2 \theta_{23} &= 0.0198\;, & 
	\frac{m_u}{m_c} &= 0.704\;,     & \frac{m_c}{m_t} &= 0.677\;, 
	\notag \\
	\sin^2 \theta_{13} &= 1.86 \times 10^{-4}\;, & \delta_{\CP} &= 0.9998\pi\;,  & \frac{m_d}{m_s} &= 0.658\;,                  & \frac{m_s}{m_b} &= 0.687\;.
\end{align}
These results correspond to a $\chi^2 \sim 3.4 \times10^7$. 
The value of $\vev{\tau}$ at the minimum is $4.636 \times 10^{-5} + 1.155\ii$, which is in the vicinity of $\tau=\ii$.

For model IIA with a $2T$ modular flavor symmetry, fitting the model to the data is not possible. 
At the minimum of the $\chi^2$, the relevant observables take the values
\begin{align}
	\sin^2 \theta_{12} &= 1.607\times 10^{-5}\;, &\sin^2 \theta_{23} &= 3.912 \times 10^{-7}\;, &
	\frac{m_u}{m_c} &= 0.0023 \;, &\frac{m_c}{m_t} &= 0.0025\;,
	\notag \\
	\sin^2 \theta_{13} &= 8.270 \times 10^{-6}\;,  &\delta_{\CP} &= 0.185\pi\;, 
	&\frac{m_d}{m_s} &= 0.7047\;,  &\frac{m_s}{m_b} &= 0.5866\;.
\end{align}
These lead to a global $\chi^2 \sim 3.6 \times 10^5$. 
At the minimum, we have $\vev{\tau} = -0.307 + 3.804\ii$, located close to $\tau\to\ii\infty$, as expected from the near-critical analysis.

For model IIIA based on $\Gamma(2)/\Gamma(6)$, it is possible to fit the model to observations with excellent accuracy. 
To some extent, this is a consequence of the presence of a large number (21) of free parameters available in this model.
At the minimum of $\chi^2$, the predicted observable values are
\begin{align}
	\sin^2 \theta_{12} &= 0.0508\;, & \sin^2 \theta_{23} &= 0.00160\;, & 
	\frac{m_u}{m_c} &= 0.00193\;, & \frac{m_c}{m_t} &= 0.00280\;, 
	\notag \\
	\sin^2 \theta_{13} &= 1.21 \times 10^{-5}\;,  & \delta_{\CP} &= 0.384\pi\;, &\frac{m_d}{m_s} &= 0.0505\;, &\frac{m_s}{m_b} &= 0.0182\;.
\end{align}
All of the observables fall within the 1$\sigma$ range. The value of $\vev{\tau}$ at the minimum is $-0.285 + 2.995\ii$, located in the vicinity of $\tau\to\ii\infty$.

Let us now move on to the next class of model, as presented in \Cref{tab:ModelsummaryB}.
For convenience, we provide explicitly the representation and weight assignments here
\begin{subequations}\label{eq:Model12RepWeight}
\begin{align}
\rho^{(u^c)} &: \rep{2}'' \oplus \rep{1}\;, \quad & k_{u^c_{1,2,3}} &= (1,1,2)\;,\quad  &\rho^{(d^c)}     &: \rep{3}\;, \quad  & k_{d^c}     &= 2\;, \\
\rho^{(Q)}   &: \rep{3}\;,                      & k_{Q}           &= 1\;,         &\rho^{(H_{u/d})} &: \rep{1}\;,      & k_{H_{u/d}} &= -1\;.
\end{align}
\end{subequations}
The sole difference between this class of models and the previous one lies in the weight assignment for the up-quark sector, which now is $k_{u^c_{1,2,3}} = (1, 1, 2)$ instead of $k_{u^c_{1,2,3}} = (3, 3, 2)$.

We have seen that models IA and IIA, based on $\Gamma_3'$ and $2T$ respectively, cannot yield realistic phenomenology. Similarly, the choices in \Cref{eq:Model12RepWeight} do not lead to phenomenologically viable results. In particular, in model IB with modular flavor symmetry $\Gamma_3'$, we see that the up-quark mass matrix is singular, cf.~\Cref{tab:ModelsummaryB}, leading to the unrealistic result $m_u/m_c=0$. Further, model IIB, based on the group $2T$, cannot generate two of the up-type quark masses because, as can be read off from \Cref{tab:ModelsummaryB}, the up-quark mass matrix has only rank one in this case.

In contrast, our model IIIB is endowed with a $\Gamma(2)/\Gamma(6)$ symmetry.
In this case, as for model IIIA, it is possible to fit the model to the experimental data with remarkable accuracy, as the number of free parameters exceeds the number of observables. 
At the minimum of the $\chi^2$, we find
\begin{align}
	\sin^2 \theta_{12} &= 0.0525\;, & \sin^2 \theta_{23} &= 0.00156\;, & \frac{m_u}{m_c} &= 0.00207\;,   & \frac{m_c}{m_t}    &= 0.00274\;, \notag \\
	\sin^2 \theta_{13} &= 1.41 \times 10^{-5}\;,& \delta_{\CP}    &= 0.366\pi\;,
	 & \frac{m_d}{m_s}   &= 0.0501\;,             & \frac{m_s}{m_b} &= 0.0181\;.
\end{align}
All of the observables are within the $1\sigma$ range. 
The modulus value at the minimum is $\vev{\tau} = -0.139 + 2.991\ii$, in the vicinity of $\tau\to\ii \infty$.

Let us make a final remark on models furnished with a large number of free parameters. 
Like the models IIIA and IIIB presented here, constructions with many parameters can fit the experimental data very well, but their predictive power is limited. 
However, some of the free parameters do not represent true degrees of freedom.
For example, in the $\Gamma(2)/\Gamma(6)$ model IIIA, the term $\mathcal{Y}^{(3)}_{\rep{2'}} u^c_D Q H_u \subset \mathscr W$, where
\begin{equation}
\label{eq:G2RG6Y2pk3}
	\mathcal Y_{\rep{2'}}^{(3)} = \alpha_4 Y_{\rep{2'}I}^{(3)} + \alpha_5 Y_{\rep{2'}{II}}^{(3)} + \alpha_6 Y_{\rep{2'}{III}}^{(3)} + \alpha_7 Y_{\rep{2'}{IV}}^{(3)}
\end{equation}
has in principle four free parameters. 
Once $\tau$ settles at its \ac{VEV}, the true degrees of freedom of $\mathcal Y_{\rep{2'}}^{(3)}$, being a two-dimensional object, must be only two. 
Thus, two of the free parameters are, in fact, redundant. Hence, the actual number of true degrees of freedom in the model is below the number of parameters. 
In general, for each superpotential term with a generic modular form, the degrees of freedom are limited by the dimension of the required \ac{VVMF}.

\end{appendix}

\clearpage
\bibliographystyle{OurBibTeX}
\bibliography{references}
\end{document}